\documentclass[12pt,preprint]{aastex}
\usepackage{graphicx}
\usepackage{placeins}
\newcounter{subfigure}

\begin{document}

\title{	Dusty Winds: Extraplanar PAH Features of Nearby Galaxies}
	
\author{Alexander McCormick\altaffilmark{1}, Sylvain Veilleux\altaffilmark{1,2,3,4}, and David S. N. Rupke\altaffilmark{5}}

\altaffiltext{1}{Department of Astronomy, University of Maryland, College Park, MD 20742, USA; E-mail: alexm@astro.umd.edu, veilleux@astro.umd.edu}

\altaffiltext{2}{Joint Space-Science Institute, University of Maryland,
 College Park, MD 20742, USA}

\altaffiltext{3}{Astroparticle Physics Laboratory, NASA Goddard Space
 Flight Center, Greenbelt, MD 20771, USA}

\altaffiltext{4}{Max-Planck-Institut f\"ur extraterrestrische
 Physik, Postfach 1312, D-85741 Garching, Germany}

\altaffiltext{5}{Rhodes College, 2000 N. Parkway, Memphis, TN 38112, USA; E-mail: rupked@rhodes.edu}

\begin{abstract}
Recent observations have shown the presence of dust and molecular material in galactic winds, but relatively little is known about the distribution of these outflow components. To shed some light on this issue, we have used IRAC images from the {\em Spitzer Space Telescope} archive to investigate polycyclic aromatic hydrocarbon (PAH) emission from a sample of 16 local galaxies with known winds. Our focus on nearby sources (median distance 8.6 Mpc) has revealed detailed PAH structure in the winds and allowed us to measure extraplanar PAH emission. We have identified extraplanar PAH features on scales of $\sim$ 0.8 - 6.0 kiloparsecs. We find a nearly linear correlation between the amount of extraplanar PAH emission and the total infrared flux, a proxy for star formation activity in the disk. Our results also indicate a correlation between the height of extraplanar PAH emission and star formation rate surface density, which supports the idea of a surface density threshold on the energy or momentum injection rate for producing detectable extraplanar wind material.
\end{abstract}


% ----- INTRODUCTION -----

\section{INTRODUCTION}
\label{intro}
Superwinds are galaxy-scale outflows of material often present in galaxies hosting star formation or active galactic nuclei (AGNs) \citep{hec90,leh95,vei05}. The feedback from winds plays an important role in the evolution of galaxies, their interstellar media (ISM), and the surrounding intergalactic or intracluster medium in which they are embedded. Outflows of material can inhibit the growth of a central supermassive black hole and limit the star formation rate (SFR) by removing the fuel for star formation \citep{dim05,nar08,hop10}.

Prior to 2005, much of the observational data have emphasized the entrained gas in winds from the neutral gas \citep{hec00,rup02,rup05a,rup05b,rup05c,sch04,mar05} to ionized gas \citep{hec90,leh95} to the highly ionized X-ray emitting plasma \citep{rea97,pie00,mcd03,ehl04,huo04,str04a,str04b}. Recent observations have shown that these outflows entrain dust \citep{hec00,tac05,eng06,rou10} and molecular gas \citep{vei09,fis10,fer10,irw11,stu11,ala11,aal11}.

The striking detection by \cite{eng06} of polycyclic aromatic hydrocarbon (PAH) emission in the prototypical wind of M82 has demonstrated that the four near-infrared (NIR) to mid-infrared (MIR) channels of the InfraRed Array Camera (IRAC) on board the {\em Spitzer Space Telescope} are ideal for detecting warm dust and molecules in galactic winds. PAHs are aromatic molecules a few \AA \ in diameter and are often referred to as small dust grains. Tens of carbon atoms with a "fringe" of hydrogen atoms make up a typical PAH molecule, while the largest ones contain hundreds of carbon atoms. These molecules are found throughout the ISM of our Galaxy as well as in nearby galaxies exhibiting evidence of ongoing or recent star formation. Since they efficiently absorb ultraviolet and optical photons, their reprocessed infrared (IR) emission features are often much stronger than the background dust continuum \citep{smi07}. PAHs are widely believed to be responsible for emission features in the NIR to MIR at 3.3, 6.2, 7.7, 8.6-8.8 and 11.3-11.9 $\mu$m \citep{leg84,all85,all89}.

The 3.6, 4.5 and 5.8 $\mu$m IRAC channels can detect dust at temperatures a few hundred to 1000 K, and the 3.6, 5.8 and 8.0 $\mu$m channels are sensitive to PAH emission features. The 8.0 $\mu$m band is particularly important due to the 7.7 $\mu$m PAH feature (e.g. Figure 1 of \citealt{rea06}) which typically dominates compared to other flux contributions from stars, dust continuum, molecular emission, and fine structure lines (see discussion in \S~\ref{convolution}). Relative to the 8.0 $\mu$m band, the 4.5 $\mu$m band contains very little PAH emission while significantly more stellar and dust continuum emission. Some H$_2$ rotational line emission and trace Br$\alpha$ emission may also contribute in the 4.5 $\mu$m band \citep{eng06}. The 3.6 and 5.8 $\mu$m bands also contain the 3.3 and 6.2 $\mu$m PAH emission features respectively, so using these bands as tracers of stellar emission is less reliable. Therefore, a comparison of 4.5 and 8.0 $\mu$m images differentiates PAH emission from stellar emission. We have used this comparison to identify extraplanar PAH features and estimate extraplanar PAH flux. Our sample of local galaxies provides the resolution necessary to distinguish coplanar from extraplanar emission and to study morphological features. In this paper, PAH emission or PAH flux refers to measurements in the IRAC 8.0 $\mu$m band rather than a bolometric PAH measurement.

In \S~\ref{sample}, we describe the selection criteria for our sample of nearby galaxies and provide tables summarizing their basic properties and the data we used from the {\em Spitzer} archive. In \S~\ref{reduce} and~\ref{analysis}, we describe the data reduction process with example images and our analysis respectively. We present our results in \S~\ref{results_discuss}, include images of each galaxy in the IRAC 4.5 and 8.0 $\mu$m channels, discuss some of the most prominent features we have identified, compare our results with previous observations, and describe variations with galaxy type, IR luminosity, and SFR surface density. Our results are summarized in \S~\ref{summary}. Appendix~\ref{append:galaxies} contains discussion of each individual galaxy and its features.

% ----- SAMPLE -----

\section{SAMPLE}
\label{sample}

We selected the 16 galaxies in our sample (listed in Table~\ref{tbl:sample}) based on three main criteria. (1) From observations at other wavelengths, they are known to host galactic-scale winds or exhibit associated extraplanar material. Most have disk morphology with inclinations ($i$) typically $\sim$ 70$^\circ$, making extraplanar regions distinguishable from the galaxy's disk. (2) They are nearby galaxies with a median distance of 8.6 Mpc. We adopted redshift-independent distance values (except for NGC 2992) found using the tip of the red giant branch or the Tully-Fisher relation. By choosing local sources, we maintain good spatial resolution even for the most distant galaxy in our sample: at the adopted distance of 31 Mpc for NGC 2992, the 1.72" and 1.98" mean FWHM of the 4.5 and 8.0 $\mu$m IRAC bands resolve structure down to  $\sim$ 260 pc and $\sim$ 300 pc respectively. (3) Each source has publicly available IRAC data in the {\em Spitzer} archive (summarized in Table~\ref{tbl:data}). Data for four of the galaxies (NGC 1482, NGC 1705, NGC 4631, and M82) came from the SINGS survey \citep{ken03}.

From previous multi-wavelength observations, most of the galaxies host active star formation (H {\sc ii} regions), two (NGC 3079 and NGC 4945) contain both H {\sc ii} regions and an AGN and two (NGC 2992 and NGC 4388) exhibit AGN but little or no ongoing star formation. Since NGC 55 and NGC 891 are not generally considered starburst galaxies, but do contain active star forming regions, we used the broader catch-all 'H {\sc ii}' type to include a range of SFRs.

\begin{figure}[htbp]
\centering
\includegraphics[width=1.0\textwidth]{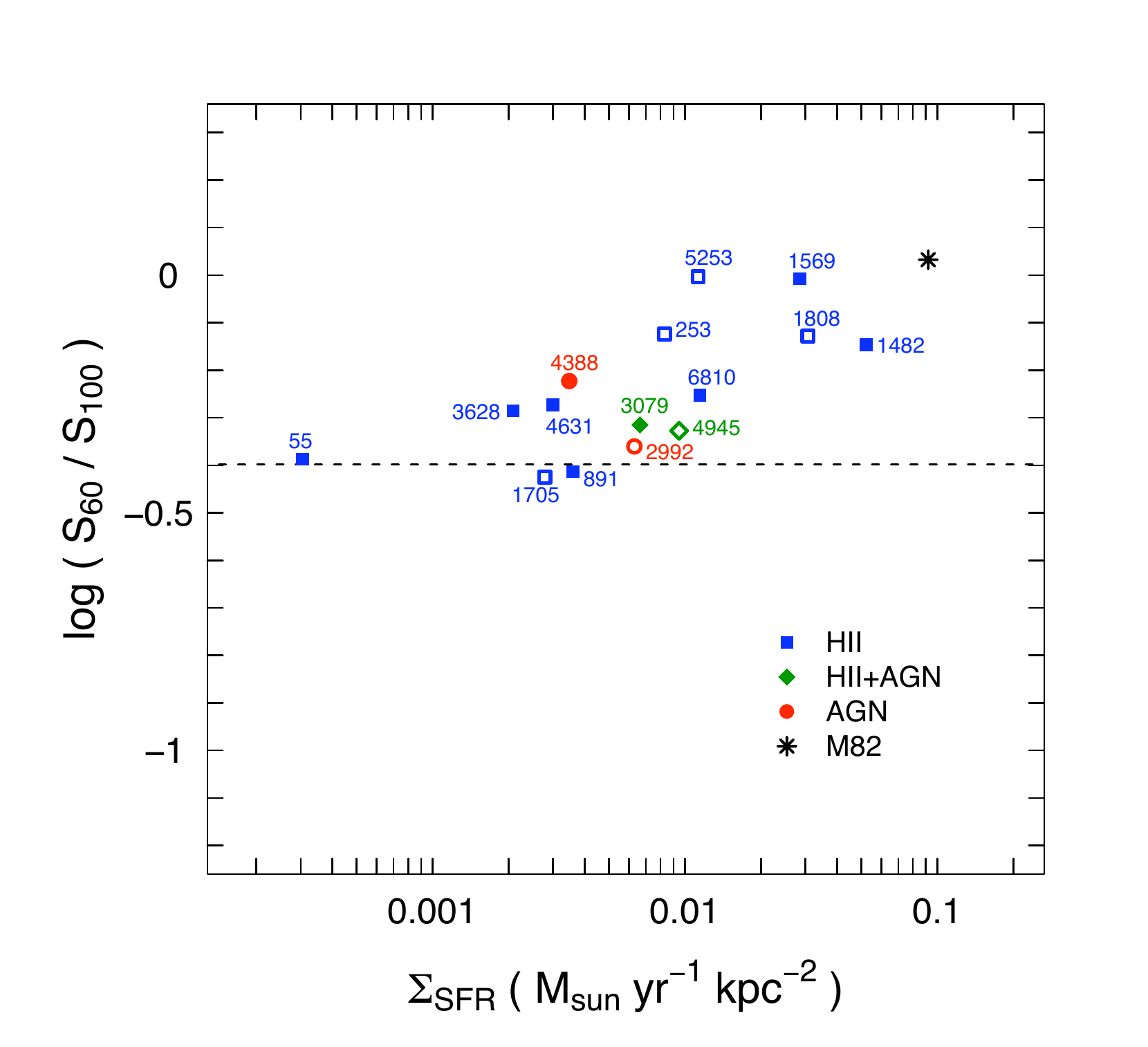}
\caption{\footnotesize{Far-infrared color vs. SFR surface density as analog to Fig. 5 in \cite{ros03}. We have included their threshold for IRAS warm galaxies at $S_{60}$/$S_{100}$ $\geq$ 0.4 as the dashed line. $\Sigma_{SFR}$ was derived from \cite{ken98} using the $D_{25}$ values in Table~\ref{tbl:sample} for consistency with the method of \cite{ros03}. Open points represent galaxies for which we used a 20-$\sigma$ region to define the disk in our flux measurements, while solid points used a fitted scale height region (see \S~\ref{flux}). Points are labeled with each galaxy's NGC number (except for M82).}}
\label{fig:RossaDettmar_analog}
\end{figure}

\cite{ros03} investigated extraplanar diffuse ionized gas (eDIG) in H$\alpha$ for a sample of nearby edge-on spiral galaxies. Their sample covered a range of IR luminosities ($L_{IR}$) in order to determine minimum thresholds for SFR surface density ($\Sigma_{SFR}$ - from far-IR surface brightness, $L_{FIR}$/$D_{25}^2$) and far-IR flux ratio ($S_{60}$/$S_{100}$) leading to the presence of eDIG. All the galaxies in our sample apart from NGC 55 lie above their derived lower limit for $L_{FIR}$/$D_{25}^2$ ((3.2 $\pm$ 0.5) $\times$ 10$^{40}$ erg s$^{-1}$ kpc$^{-2}$) and lie either near or above their limit for $S_{60}$/$S_{100}$ (0.4). Combining their data with that of \cite{leh95}, \cite{ros03} noted the apparent correlation between $\Sigma_{SFR}$ and far-IR flux ratio, which is also recovered in our sample as shown in Figure~\ref{fig:RossaDettmar_analog}. The $\Sigma_{SFR}$ shown in Figure~\ref{fig:RossaDettmar_analog} was derived from \cite{ken98} using the $L_{IR}$ and $D_{25}$ values in Table~\ref{tbl:sample}.

% ---- DATA REDUCTION -----

\section{DATA REDUCTION}
\label{reduce}
Extended, extraplanar emission features are typically faint compared to their host galaxy and its nuclear regions. These features are particularly susceptible to confusion with data artifacts even for nearby, resolved galaxies. Therefore, a careful data reduction process is necessary to distinguish emission features from a number of potential artifacts.

Our sample contains data from a variety of observations, and each of the galaxies has different properties and morphology, so the lack of uniformity made it impossible to use a single data reduction rubric for all sources. Therefore, we tailored our approach to each data set's individual features and artifacts. Typically, we used the basic calibrated data, uncertainty and mask FITS files from the {\em Spitzer} archive as our starting point. Since we were most interested in the extraplanar emission, maintaining the quality of the galaxy's nuclear region was sometimes sacrificed in order to get a better result for extraplanar regions. We also dealt with other features such as cosmic ray hits, image alignment, and rotation. However, we dealt with these relatively minor issues in standard ways, so we do not discuss them here.

\subsection{Removal of IRAC Artifacts}
\label{IRAC}
The mitigation of artifacts in the IRAC basic calibrated data presents a variety of challenges due to electronic, instrumentation, and environmental effects. Here, we discuss the major artifact corrections performed on the data in our sample. A comprehensive discussion of the IRAC pipeline and data artifacts can be found in the IRAC instrument handbook\footnote{http://ssc.spitzer.caltech.edu/irac/iracinstrumenthandbook/}.

\subsubsection{SSC-contributed Artifact Correction Code}
\label{SSC}
All of the IRAC data sets from the {\em Spitzer} archive contain basic calibrated data images (BCDs), uncertainty maps, and mask files amongst their data products. These products are the result of the {\em Spitzer Science Center} (SSC) pipeline which corrects for detector bias, does dark subtraction, flatfielding and photometric calibrations. Prior to release, some of the data were also run through a SSC-contributed artifact mitigation code\footnote{http://ssc.spitzer.caltech.edu/dataanalysistools/tools/contributed/irac/iracartifact/}, so these data contain corrected basic calibrated data (CBCDs) in addition to the original BCDs. The SSC-contributed code attempts to correct for the following artifacts:

\begin{itemize}
\item Column pulldown/pullup - a uniform bias shift up or down within columns containing a bright source. According to the IRAC instrument handbook, this artifact only appears in the 3.6 and 4.5 $\mu$m channels, but we also observed a similar artifact in some 8.0 $\mu$m data (see \S~\ref{banding}).
\item Muxbleed - a periodic shift in the bias along rows in the image containing a bright source such as a star. It appears like a decaying "trail of breadcrumbs". This artifact only affects the 3.6 and 4.5 $\mu$m channels.
\item Electronic banding - a nearly constant bias shift to higher values across rows containing a bright source. This artifact appears only in the 5.8 and 8.0 $\mu$m channels.
\item First frame effect - a bias effect on the entire frame, which makes consecutive frames mismatched - a problem when co-adding images or creating mosaics. In our mosaics and image stacks, we excluded the first short frame taken at each new pointing, since those frames are most affected by the first frame effect. 
\end{itemize}

The SSC-contributed code does not correct full-array pull-up, internal scattering, optical banding, "muxstriping", or persistent images. For a detailed description of these artifacts, consult the IRAC instrument handbook. We discuss the correction of full-array pull-up, internal scattering, and optical banding in the following sections. The data are minimally affected by "muxstriping" or persistent images, so in general these artifacts could be ignored. Whenever advantageous, we worked from the CBCDs, since they often exhibited a useful reduction or elimination of the prescribed artifacts. As implied above, some data sets only contained BCDs and no CBCDs. For those data sets, we downloaded the SSC-contributed code and used it to produce CBCDs.

\subsubsection{Full-Array Pull-Up and Internal Scattering}
\label{background}
When integrating individual frames in both the 4.5 $\mu$m and 8.0 $\mu$m channels, the total flux from all sources within the frame contributes a uniform, proportional offset across all pixels - an effect called "full-array pull-up" by the IRAC instrument handbook. In addition to this effect, some light can scatter approximately uniformly within the array due to the presence of a bright source in the 8.0 $\mu$m channel, adding to the total offset. We accounted for the offsets due to full-array pull-up and internal scattering by background subtracting all individual frames via user-defined background regions. The background regions were carefully selected to exclude galaxy features, foreground stars, and any data artifacts. Bright features within the galaxy's disk or any bright stars within the frame were the main source of the internally scattered flux, but these features were not part of our extraplanar flux measurements, so removing their redistributed flux from the frame was an acceptable solution.

Background matching via subtraction worked significantly better than utilizing the MOPEX Overlap modules, which tended to introduce a bias across the mosaic. It does not escape our notice that any uniform extraplanar emission would be subtracted off using this method, but we expect such a flux distribution to be unlikely (a test of this background subtraction method is described in \S~\ref{model_ePAH} below).

\subsubsection{Electronic and Optical Banding}
\label{banding}
The horizontal (row) electronic and optical banding effect was pervasive in the data and presented a significant challenge in exposures where a galaxy covered a large fraction of the array or in exposures where banding and extraplanar features overlapped. In the course of inspecting the CBCDs, we found that the banding correction applied to the 8.0 $\mu$m data by the SSC-contributed code and later versions of the SSC pipeline provided incomplete artifact removal by only addressing the brightest rows. 

\begin{figure}[htbp]
\centering
\includegraphics[width=1.0\textwidth]{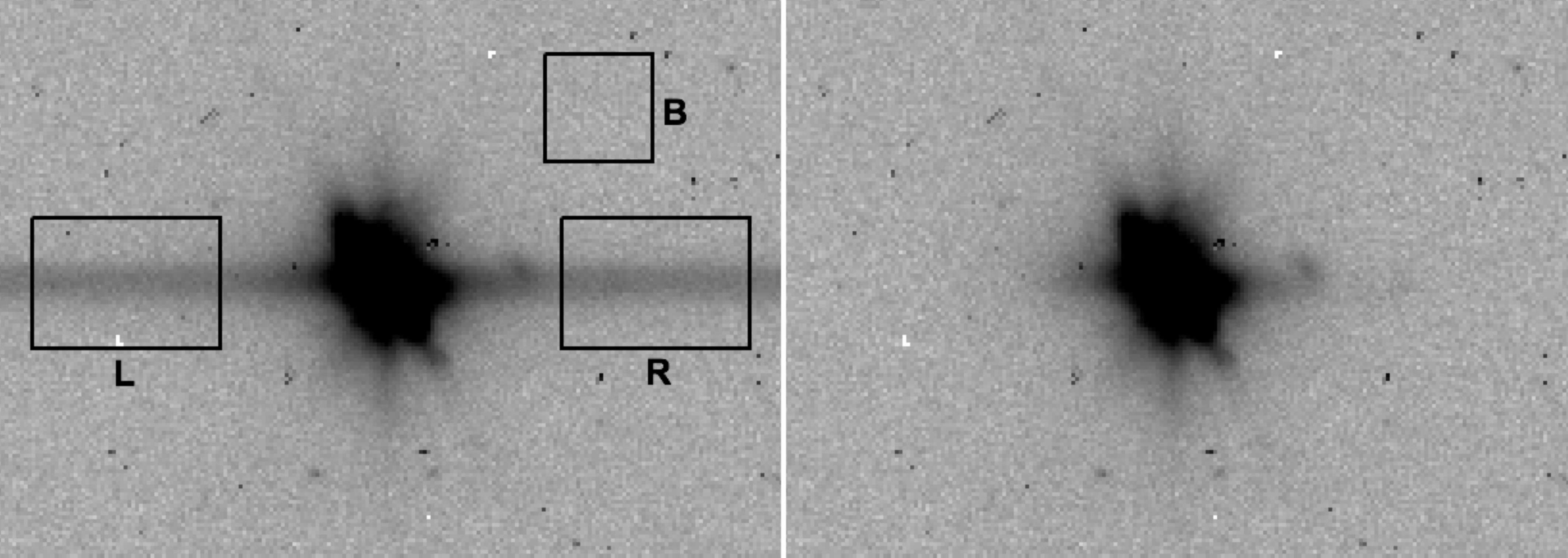}
\caption{\footnotesize{Before (left) and after (right) electronic banding correction using our PyRAF script on a single IRAC 8.0 $\mu$m frame of NGC 1482. The regions labeled L and R sample the areas affected by the banding artifact and define the rows where the subtraction takes place. The region labeled B samples the background.}}
\label{fig:banding}
\end{figure}

Since the banding artifacts for extended sources covered a wide range of rows, we developed a PyRAF script to correct individual frames by removing the entire band rather than just the few brightest rows. Our PyRAF script employs user-defined rectangular regions to sample the background and banding artifact regions. It then subtracts a constant value banding offset from each affected row. Overlaid example regions are shown in the left panel of Figure~\ref{fig:banding} and an example result is shown in the right panel.

We also observed a less common, but similarly uniform bias shift in the columns of 8.0 $\mu$m images containing a bright source. However, in this case the bias shift was uniformly negative rather than positive. We adapted our PyRAF script and corrected the affected columns.

\subsubsection{PSF Fitting and Subtraction}
\label{prf}
The point spread function (PSF) wings of bright sources can be another locus of confusion with extraplanar emission. Although the galaxies in our sample are not true point sources, several (NGC 1482, NGC 2992, NGC 3079, NGC 3628 and NGC 5253) contain bright, concentrated nuclei, which exhibit PSFs very similar to those produced by bright stars and extend to regions of possible extraplanar emission. In addition, several galaxies (e.g. NGC 1569) have bright foreground stars with prominent PSF wings in close proximity to or overlapping the galaxy's disk. The SSC has produced point response functions (PRFs), which are tables for point source fitting stored as 2-dimensional FITS files and over-sampled to account for intra-pixel sensitivity variations. By sampling the PRFs at the appropriate resolution, we fit and subtracted an estimate of the PSF wings for nuclei and foreground stars using the APEX and APEX QA modules of the SSC-provided MOPEX software.

\begin{figure}[htbp]
\centering
\includegraphics[width=1.0\textwidth]{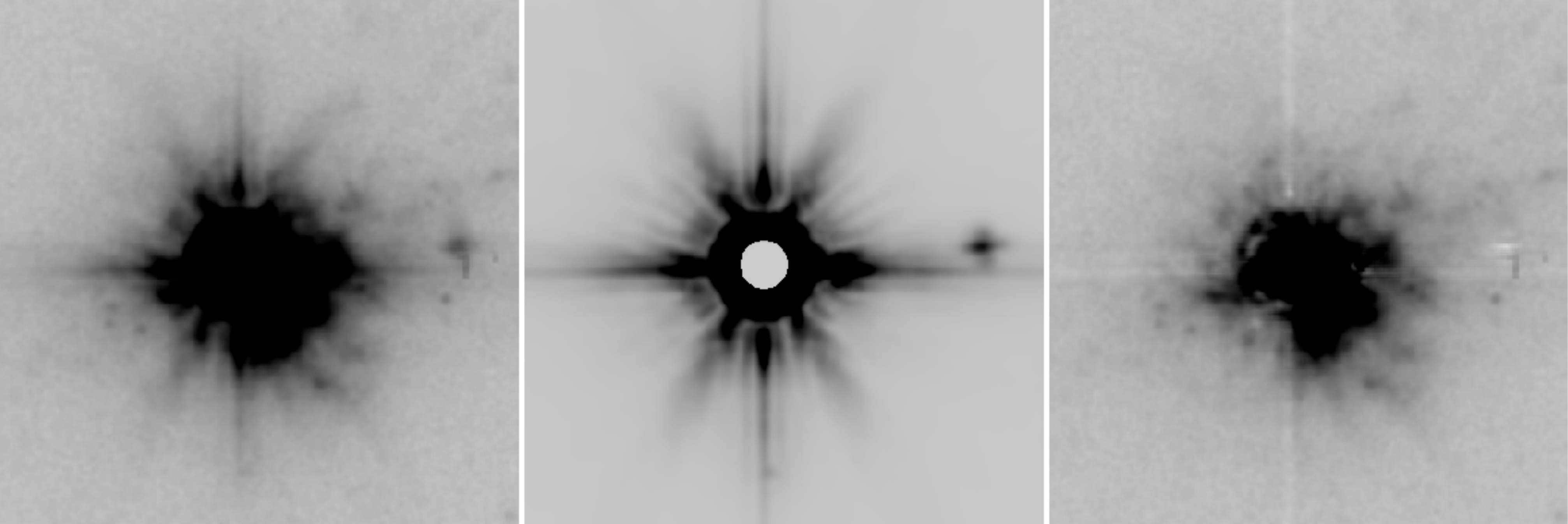}
\caption{\footnotesize{PSF fitting and subtraction for NGC 5253 (all IRAC 8.0 $\mu$m). The left panel shows the co-added and resampled image. The center panel contains the PSF fitted to the  wings of the 8.0 $\mu$m image using the SSC-produced extended PRF. The right panel shows the residual image after PSF subtraction (left - center = right). Each panel contains only the central region of these images aligned with their xy-coordinates. The left and right panels are shown on the same scale with a logarithmic stretch, but the center image has a different scale, so it has been clipped and stretched to approximately match the other images. The central region of the fitted PSF was excluded during fitting. In this example, the PSF has been somewhat over-subtracted in certain areas, as shown by the white cross-hairs pattern visible in the right panel. However, other areas seem under-subtracted, such as the area encircling the nucleus. NGC 5253 is not typical of the objects in our sample, but rather illustrates the worst case scenario where the intensity of the PSF is comparable to and overlapping with extraplanar PAH emission. Finally, in the PSF-subtracted image, we can clearly identify extraplanar emission plumes extending toward the lower right corner of the image.}}
\label{fig:PRF}
\end{figure}

Since the nuclei of our sample galaxies are not true point sources (the closest being NGC 5253), fitting their wings presented more of a challenge than foreground stars. The bright source generating the PSF is typically spread over several pixels, each with their own intra-pixel variation, leading to smeared wings, which could not be fit well for a single frame with the SSC-provided PRFs. When multiple frames were stacked or mosaicked, that reduced some of the intra-pixel variation, making a fit a bit more straightforward, but we still needed to use a combination of the APEX fitting modules in MOPEX together with some common sense "manual" adjustments, which sometimes meant a segmented model of the source to imitate the smearing. Figure~\ref{fig:PRF} shows an example of the fitting challenges.

\subsection{Aperture Corrections}
\label{aperture}
The IRAC instrument handbook lists extended source photometrical correction factors\footnote{http://irsa.ipac.caltech.edu/data/SPITZER/docs/irac/iracinstrumenthandbook/30/} based on the radius of an effective circular aperture for the flux measurement region. Since our sources are nearby, and all our extraplanar regions cover extended solid angles ($>$ 35"), we applied the infinite, asymptotic correction coefficients: 4.5 $\mu$m $\rightarrow$ 0.94 and 8.0 $\mu$m $\rightarrow$ 0.740. In order to save a step in our flux measurements, we applied the aperture corrections to the 4.5 and 8.0 $\mu$m micron images before the stellar continuum emission subtraction in the next section. Despite these aperture corrections, our flux measurements still have uncertainties at the 10-20\% level due to differences in "full-array pull-up" and internal scattering from source to source. We propagated these uncertainties through our calculations.

\subsection{Convolution, Stellar Scaling and Subtraction}
\label{convolution}
Apart from the 7.7 and 8.6-8.8 PAH emission features, other potential contributions to the 8.0 $\mu$m band flux include stellar and dust continuum emission, [Ar {\sc ii}] and [Ar {\sc iii}] fine structure lines at 6.985 $\mu$m and 8.991 $\mu$m respectively, and the H$_2$ $S$(5) and H$_2$ $S$(4) pure rotational lines at 6.909 $\mu$m and 8.026 $\mu$m respectively. \cite{smi07} included the fine structure lines and H$_2$ rotational lines in their detailed decomposition of IR spectra from 59 nearby galaxies with a range of star formation properties, but these lines typically accounted for negligible flux in the 8.0 $\mu$m band compared with the contributions of the PAH emission features plus stellar and dust continuum emission. Likewise, these atomic and molecular lines appear comparatively weaker in the nuclei of starburst galaxies \citep{bra06} as well as within ultraluminous infrared galaxies \citep{arm07,vei09ulirg}. In addition, absorption from amorphous silicate centered at 9.7-9.8 $\mu$m can attenuate PAH emission at the red end of the 8.0 $\mu$m band, particularly the 8.6-8.8 $\mu$m PAH feature \citep{bra06,smi07}. However, the optical depth due to silicate absorption will likely approach zero for extraplanar regions, so we considered it negligible. Similarly, dust continuum emission should be less important in extraplanar regions. Therefore, we focused on accounting for the stellar emission contribution to the 8.0 $\mu$m band flux. 

First, we convolved the 4.5 $\mu$m images to the resolution of the 8.0 $\mu$m images using the appropriate convolution kernel as described in \cite{gor08} along with the associated IDL routines\footnote{http://dirty.as.arizona.edu/$\sim$kgordon/mips/conv$\_$psfs/conv$\_$psfs.html}. The convolved images and original 8.0 $\mu$m images both have a PSF FWHM of $\sim$ 1.9", which is comparable to the 1.22" pixel scale of the IRAC images. As a check, we also generated kernels from the data directly, producing similar flux values after convolution. However, we used the images convolved with the kernels from K. Gordon for our stellar subtraction, since the data-generated kernels tended to over-reduce resolution and compromise the subtraction of individual stars.

\cite{dal05} used a conceptually simple approach to model the stellar contribution to the SEDs of 75 nearby galaxies. Using a Starburst99 \citep{lei99,vaz05} model with 900 Myr continous star formation, solar metallicity, and a Salpeter IMF ($\alpha_{IMF}$ = 2.35) and dust models essentially parametrized by the far-infrared color $f_{\nu}$(70 $\mu$m)/$f_{\nu}$(160 $\mu$m), they found "remarkable" fits to their sample's SEDs. With this method, they estimated the "dust-only" 8.0 and 24 $\mu$m fluxes by subtracting an extrapolated stellar contribution scaled from the 3.6 $\mu$m flux. Since the 4.5 $\mu$m band contains predominantly stellar continuum emission, it can be used along with a scaling factor to estimate the stellar contribution to the 8.0 $\mu$m band. Adopting their scaling factors, the extrapolated stellar contribution scaling factor to go from 4.5 to 8.0 $\mu$m is $\eta^{8*}_{\nu}$ = 0.352. Thus, in order to obtain fluxes with only 8.0 $\mu$m PAH emission, we subtracted the scaled 4.5 $\mu$m flux:

\begin{equation}
\label{eq:scaling}
f_{PAH} \ = \ f_{8.0 \mu m} \ - \ \eta^{8*}_{\nu} f_{4.5 \mu m}
\end{equation}

\begin{figure}[htbp]
\centering
\includegraphics[width=0.6\textwidth]{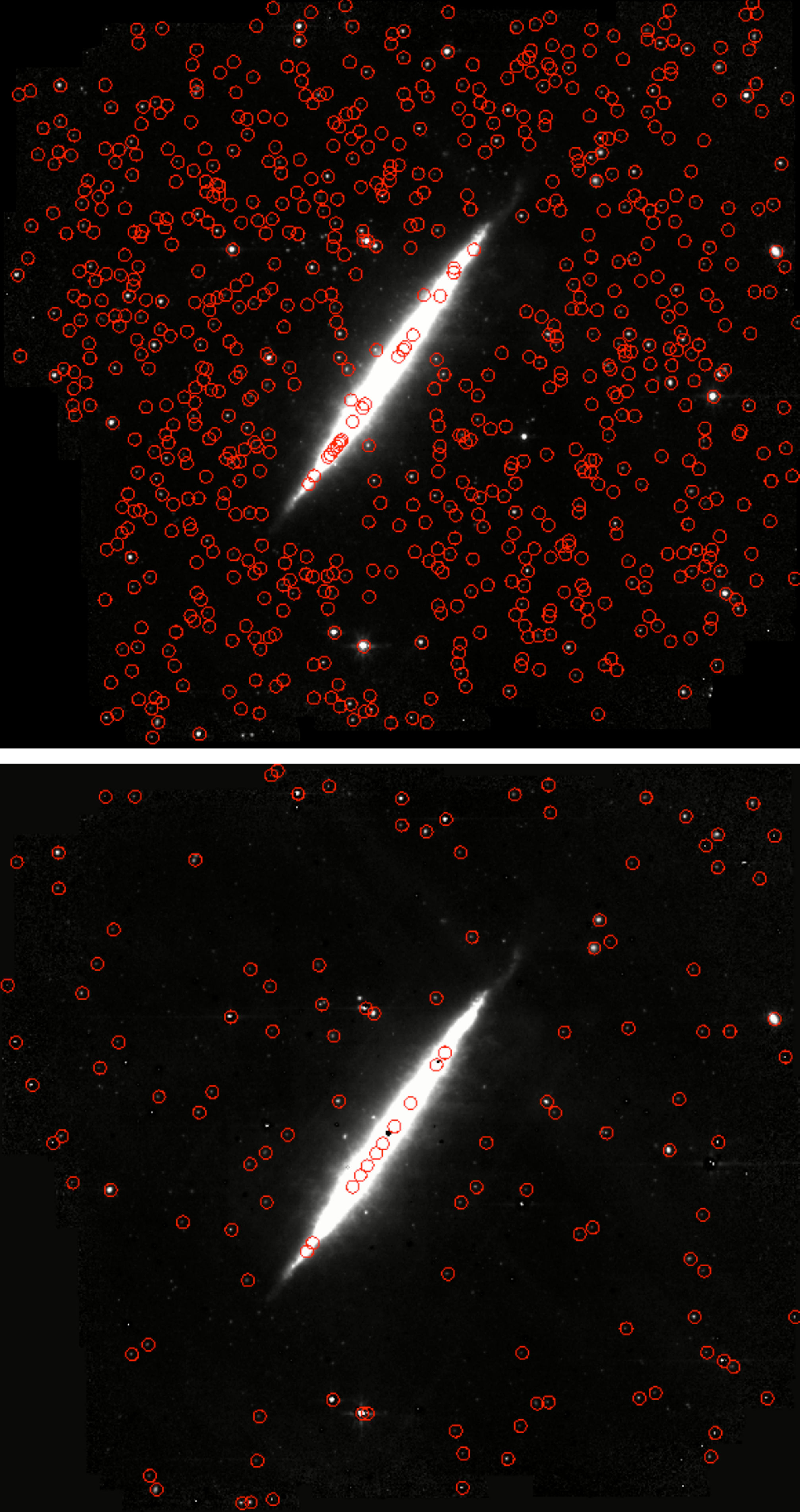}
\caption{\footnotesize{Subtraction of the stellar continuum from the data of NGC 891. The top panel shows the IRAC 8.0 $\mu$m map with overlaid red circles indicating 718 point sources (excluding those found in the galaxy's disk) detected with the APEX module in the MOPEX software. The bottom panel shows the scaled stellar continuum-subtracted map and 129 overlaid point source detections using the same APEX parameters. Mostly due to IRAC artifacts the brighter foreground stars tend not to subtract as well, so this is a systematic effect rather than a deficiency in our stellar continuum scaling.}}
\label{fig:subtract}
\end{figure}

\noindent A couple IRAC artifacts tend to redistribute the flux of bright sources, so star subtraction with this method doesn't work as consistently well for bright stars as it does for dimmer stars. When residual stars remained in the extraplanar regions of our stellar continuum-subtracted maps, we masked them out before taking flux measurements (see \S~\ref{flux}). Figure~\ref{fig:subtract} shows an example of subtracting the stellar continuum from the data for NGC 891. In subsequent sections, we refer to the stellar continuum-subtracted 8.0 $\mu$m maps as PAH images.

\subsection{Test of the Background Subtraction}
\label{model_ePAH}
Since measuring diffuse extraplanar flux depends critically on the accuracy of the background subtraction, we tested our method by carefully constructing a model of diffuse extraplanar PAH emission for NGC 891. We chose NGC 891, because (1) it has one of the best behaved edge-on disk morphologies of our sample galaxies, making it simpler to model the underlying stellar disk at 4.5 $\mu$m, (2) it is dominated by a very diffuse, nearly featureless extraplanar PAH emission morphology, and (3) it is large on the sky, with the disk subtending more than two full IRAC fields of view. Attributes (2) and (3) make quantifying the extraplanar PAH emission in NGC 891 most challenging, and thus an excellent test case for our methods of data reduction. We modeled the diffuse extraplanar PAH emission as a basic exponential disk:

\begin{equation}
\label{eq:model}
I_{PAH} \ = \ I_{0} \ * \ e^{-z / H_z} \ * \ e^{-R / H_R}
\end{equation}

\noindent where $I_{PAH}$ is the pixel intensity as a function of $z$ and $R$, $I_0$ is the reference intensity, and $H_z$ and $H_R$ are the vertical and radial scale heights with respect to the disk. We chose values of $I_0$, $H_z$, and $H_R$ which came close to replicating the morphology of the 8.0 $\mu$m data, but with more extended, diffuse flux in the $z$-direction. We chose a more extended $z$-direction morphology in order to make it harder to find background reference regions within individual on-source frames.

In order to simulate each individual frame of 8.0 $\mu$m data in the same pattern as the observations, we added (1) a scaled 4.5 $\mu$m frame, (2) the corresponding 256x256 pixel area from our model, (3) 20-30\% (the percentage was chosen at random in that range) of the total flux from the first two components evenly distributed throughout the 256x256 frame, and (4) a constant offset. Components (1) and (2) approximate the source's 8.0 $\mu$m flux, while (3) and (4) approximate the behavior of the detector, with component (3) mimicking the "full-array pull-up" plus internal scattering. The 20-30\% range was chosen to bracket the $\sim$ 25\% value listed in the IRAC instrument handbook.

Next, we performed background subtraction on the individual frames by following the procedure described in \S~\ref{IRAC}. We made a mosaic of these simulated 8.0 $\mu$m frames using MOPEX, performed scaled stellar subtraction, and applied aperture corrections, obtaining a model of diffuse extraplanar PAH emission.

We measured the flux of our modeled diffuse extraplanar PAH emission and recovered $\sim$ 98\% of the flux originally contained within the same extraplanar region of the model. This flux loss of $\sim$ 2\% is an order of magnitude lower than the $\sim$ 10-20\% systematic uncertainty of IRAC flux measurements as listed in the IRAC instrument handbook, so our method of background subtraction has a negligible effect on the measured fluxes. Since our method of background subtraction performed well in the challenging case of NGC 891, we are confident that our reduction also worked well for the other galaxies in our sample.

% ---- DATA ANALYSIS ----

\section{DATA ANALYSIS}
\label{analysis}
\subsection{Extent of Extraplanar PAH Features}
\label{z_ext}
Measuring the extent of extraplanar PAH emission features gives a basis for comparison to other data such as H$\alpha$, X-ray, Na I D, etc. In order to measure the extent of extraplanar PAH features ($z_{ext}$), we measured perpendicularly from the major axis of the disk to the tip of the feature as it appeared in the 8.0 $\mu$m images (before stellar continuum subtraction). We defined the tip of the feature by overlaying a 3-$\sigma$ contour, based on the $\sigma$ value of the background noise in the 8.0 $\mu$m image (as listed in Table~\ref{tbl:data}). The projection angles of features are unknown, so our method finds the value of $z_{ext}$ as though it were perpendicular to the line of sight setting an approximate lower limit to the actual size of these features. For cases where the galactic disks are not quite edge-on, the error in our $z_{ext}$ measurement may be larger. Table~\ref{tbl:results} lists the $z_{ext}$ value representing the most extended PAH emission feature for each galaxy.

\subsection{PAH Flux Measurements}
\label{flux}
Beyond simply identifying extraplanar PAH emission features, quantifying their flux and comparing with the host galaxy's properties allows us to investigate relationships with star formation and galaxy type. Measuring a value for the extraplanar PAH flux necessitates delineating the disk from the extraplanar region of each galaxy. Factors such as inclination and morphology can complicate this task somewhat, but our sample consists primarily of disk galaxies with nearly edge-on orientations, simplifying matters.

We wrote a PyRAF script, which determines the vertical scale height ($H_z$) of a galaxy's stellar disk and from that value generates a disk region. We used the 4.5 $\mu$m images (pre-convolution) to generate disk regions based on continuum stellar emission. Before running our script, we rotated the images to align each galactic disk horizontally. Our script first determines a vertical scale height value ($H_{col}$) for each column of pixels 'above' the disk of the galaxy by fitting a simple exponential:

\begin{equation}
\label{eq:scaleheight}
I(z) = I_0 \ e^{-(z -z_{mp})/ H_{col}} \ \ \ \ ( \ z \ \geq \ z_{mp} \ )
\end{equation}

\noindent where $I(z)$ is the intensity at height $z$,  and $I_0$ is the intensity where the fit begins. Since the vertical intensity profile plateaus near the midplane of the disk, our script omits this area by using a small offset above the midplane ($z_{mp}$) for the beginning pixel of the fit. The fit range then extends up to where the intensity has fallen off by 1/$e^2$. The script determines a global $H_z$ value by finding the median of all the $H_{col}$ values.

The actual or apparent vertical thickness of the galaxy's stellar component can vary along the disk, especially with a bulge, disk asymmetry, or inclination effects. When measuring the extraplanar flux, the biggest contaminating flux typically comes from the galaxy's bulge. Setting a constant vertical 'edge' based on just $H_z$ can include too much of the central bulge or exclude regions at larger galactic radii which should be considered extraplanar. Therefore, the method in our script for finding a vertical 'edge' ($z_{edge}$) needed to account for the bulge and other variations in disk thickness when delineating the disk region.

Starting from our assumption that {\em to first order}, the vertical light profile will have the same functional shape and $H_z$ value at all galactic radii, the exponential profile in equation~\ref{eq:scaleheight} will simply shift in the $z$-direction based on the value of $I_0$. Consider two columns which reach the same $I(z)$ at different values of $z$ - say one at large galactic radius where the projected disk intensity is low ($I_1$) and one near the nucleus where the projected disk intensity is high ($I_2$):

\begin{equation}
\label{eq:low}
I(z) = I_1 \ e^{- (z_1 - z_{mp}) / H_z}
\end{equation}
\begin{equation}
\label{eq:high}
I(z) = I_2 \ e^{- (z_2 - z_{mp}) / H_z}
\end{equation}

\noindent where $z_1$ and $z_2$ ($z_1$ \textless \ $z_2$) are the different heights at which the intensity reaches $I(z)$. If we define the shift in the $z$-direction as $z_s$ = $z_2$ - $z_1$, then

\begin{equation}
\label{eq:shift}
z_s \ = \ H_z \ln \left( \frac{I_2}{I_1} \right)
\end{equation}

\noindent Thus the shift of the exponential profile in the $z$-direction should depend only on the ratio ($I_2$ / $I_1$) and $H_z$, which we've assumed to be constant for the entire disk.

\begin{figure}[htbp]
\centering
\includegraphics[width=1.0\textwidth]{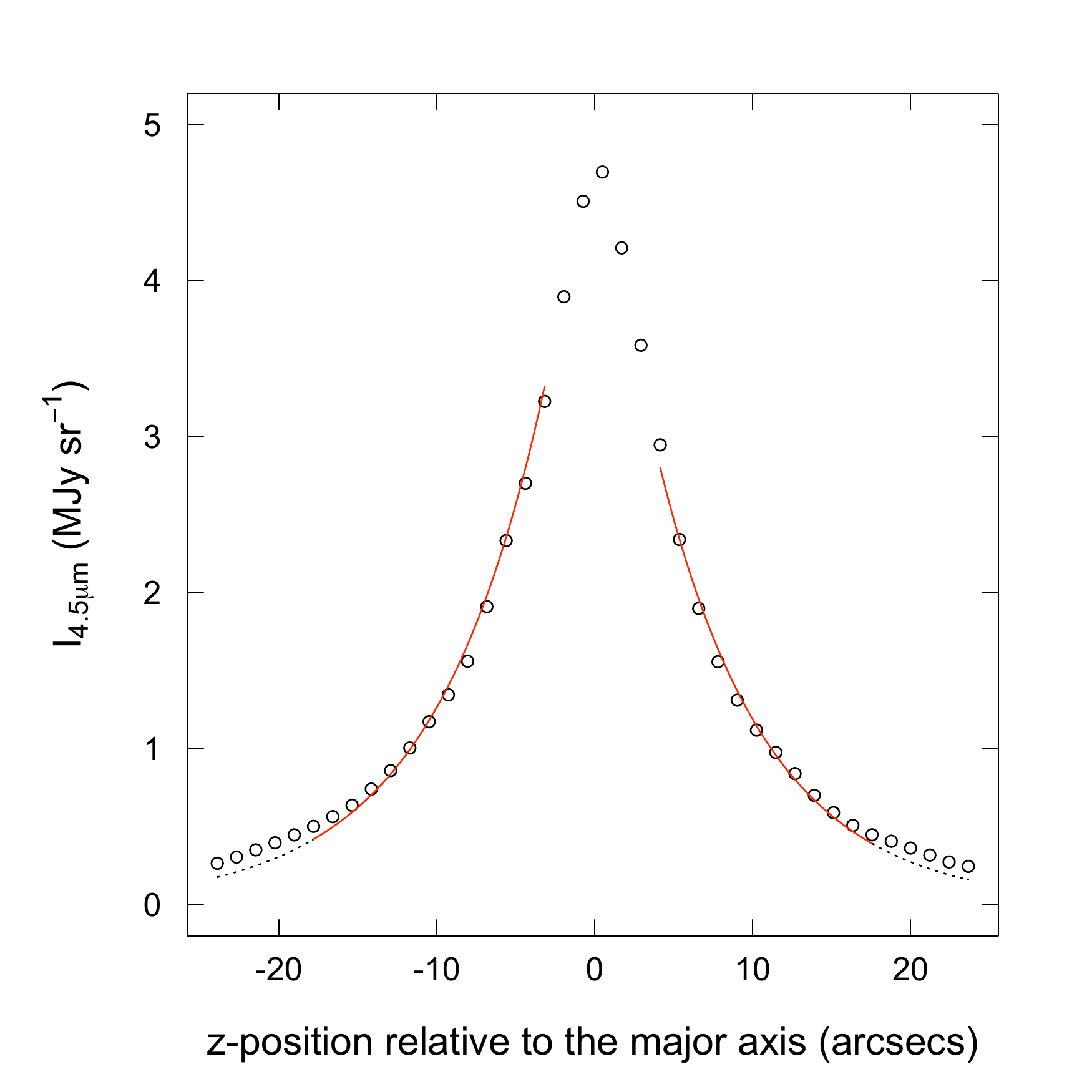}
\caption{\footnotesize{Scale height fitting to the vertical surface brightness profile (single column of pixels) of NGC 891 using the 4.5 $\mu$m data. The data are shown as black open circles. The solid red lines describe the individual exponential profiles fit to the thin disk above and below the midplane. The extent of the solid red lines indicates which pixels were used in the fits. A few central pixels near the midplane were excluded from the fits, since the profile turns over at the midplane. The dotted extension of each fit to z-positions further from the midplane illustrates the transition to the thick disk as $|z|$ increases.}}
\label{fig:H_fit}
\end{figure}

Therefore our script determines the edge of the disk region, $z_{edge}$, by combining the offset at the midplane ($z_{mp}$) with a multiple of $H_z$ and a term to account for the intensity profile shift in the $z$-direction due to a bulge or other variations in disk thickness:

\begin{equation}
\label{eq:edge}
z_{edge} = z_{mp} + 3H_z + A \ln \left( \frac{I_{bin}}{I_*} \right)
\end{equation}

\noindent where $A$ is a distance constant determined by trial, $I_{bin}$ is the median $I_0$ within a bin along the major axis of the disk, and $I_*$ is the unit intensity in MJy sr$^{-1}$. The bin size is chosen to reflect a balance of disk feature size and image resolution constraints. The factor of 3 represents a typical value for the ratio of the thick disk to the thin disk \citep[e.g.,][]{deg97}, since our method typically finds the vertical scale height of the thin disk. Including each galaxy's $H_z$ value within the shift term as suggested by equation~\ref{eq:shift} produced inconsistent results, often underestimating or overestimating the disk edge. Instead, we adopted the distance constant $A$, which we found by iteratively running our script with different constant values in order to appropriately scale the shifts of all bins. We use the same unit intensity $I_*$ as the reference value within the shift term for all galaxies in order for the $z$-direction shift to depend only on the relative values of $I_{bin}$ (i.e. the intensity variation within the disk). Figure~\ref{fig:H_fit} shows an example of vertical scale height fits to the 4.5 $\mu$m surface brightness profile above and below the midplane of NGC 891.

\begin{figure}[htbp]
\centering
\includegraphics[width=1.0\textwidth]{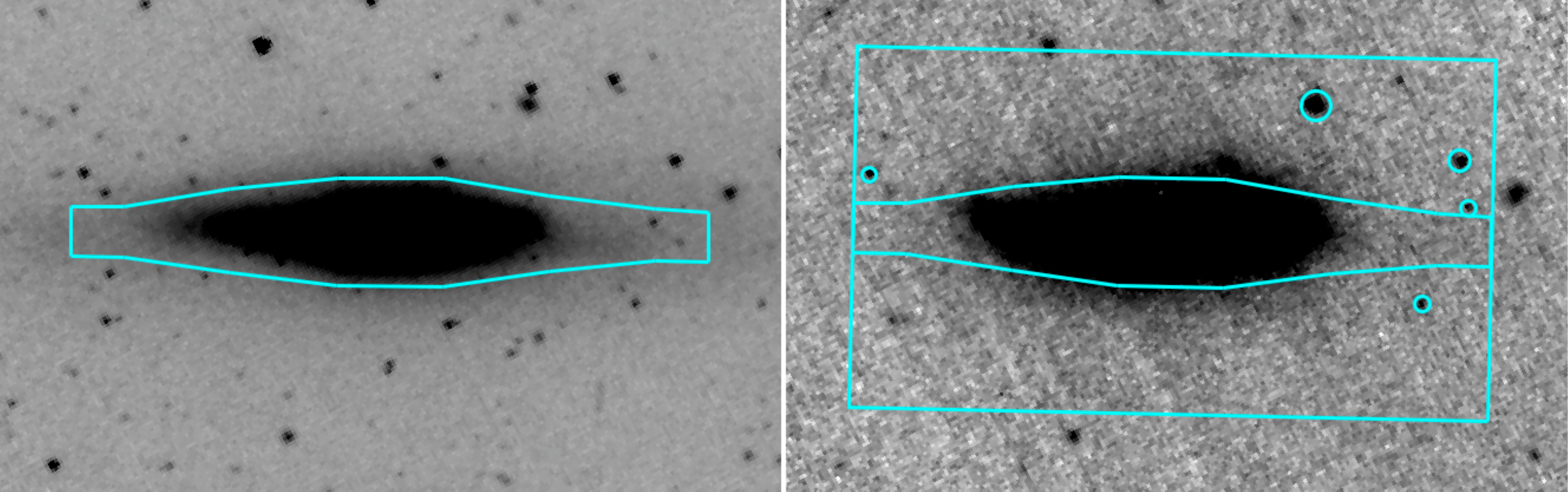}
\caption{\footnotesize{Extraplanar PAH flux calculation for NGC 4388. The left panel shows the IRAC 4.5 $\mu$m image with the disk region generated by scale height fitting. The right panel shows the IRAC 8.0 $\mu$m PAH emission image with the disk, mask and extraplanar regions overlaid. North is up, east is left in both images.}}
\label{fig:ext_flux}
\end{figure}

The midpoint of the bin and $z_{edge}$ are then the (x,y) coordinates of a point on the edge of the disk region. Once these coordinates are determined for all the bins 'above' the midplane, the script then repeats the fitting process 'below' the midplane and generates a polygonal disk region by connecting points sequentially. Our method for determining $z_{edge}$ is purposefully conservative to make sure disk regions include the thick disk and bulge, so our extraplanar flux measurements may be lower limits. Figure~\ref{fig:ext_flux} shows an example of the disk region generated with this method for NGC 4388.

For the six galaxies in our sample with lower inclination or less disk-like morphology (NGC 253, NGC 1705, NGC 1808, NGC 2992, NGC 4945, and NGC 5253), our method for finding a vertical scale height does not work. Therefore, we used a disk region defined by the 20-$\sigma$ contour of the 4.5 $\mu$m image, where $\sigma$ is the standard deviation of the background noise (see Table~\ref{tbl:data}). As a check on the validity of this method to characterize the disk region, we compared the 20-$\sigma$ contour to the disk region for galaxies where we were able to fit a vertical scale height. The results produced spatially similar regions with a typical flux difference from one region to the other of less than $\sim$ 5\%.

Overlaying the disk regions onto the PAH images, we chose rectangular regions with the same radial extent as the disk regions to enclose any extraplanar emission, but exclude features that might be due to tidal interactions (e.g. NGC 2992). Exceptions are NGC 1569, which has some extraplanar emission in a radial direction and NGC 1705, which has relatively low inclination and peculiar morphology. In several cases, artifacts or field stars remained in the PAH images, so these were masked out wherever possible. All images were background-subtracted prior to the flux calculations. Figure~\ref{fig:flux} shows all sample galaxies with disk, 20-$\sigma$, extraplanar, and mask regions overlaid.

The extraplanar PAH flux ($f_{ePAH}$) was calculated by subtracting the disk and mask regions' flux from the extraplanar region's flux for each aperture corrected 8.0 $\mu$m PAH emission image (stellar continuum-subtracted):

\begin{equation}
\label{eq:flux}
f_{ePAH} \ = \ \left[ \ \displaystyle\sum_{pixels} (I_{ext}) \ - \ \displaystyle\sum_{pixels}(I_{disk}) \ - \ \displaystyle\sum_{pixels}(I_{mask}) \ \right] d\Omega
\end{equation}

\noindent where $I_{ext}$,  $I_{disk}$ and $I_{mask}$ are the specific intensities in MJy sr$^{-1}$ of pixels in the rectangular extraplanar region, disk region, and mask region(s), respectively, and $d\Omega$ is the pixel scale in steradians. The values of $f_{ePAH}$ are listed in Table~\ref{tbl:results}.

\subsection{Characteristic PAH Properties}
\label{PAHproperties}
Quantifying the extraplanar PAH emission's characteristic height above the disk provides another avenue for comparison with galaxy properties. In order to arrive at a characteristic height, we first defined a characteristic specific intensity of the extraplanar PAH emission ($I_{ePAH}$) by finding the median pixel value of the extraplanar PAH emission (extraplanar - disk - mask) from the PAH images. We removed any potential biasing due to noise by taking the median from a subset of extraplanar emission pixels, clipped to only include pixel values 3-$\sigma$ above the background noise. Figure~\ref{fig:H_ePAH_calc} shows an example of contours at 3-$\sigma$ above the background and at the calculated value of $I_{ePAH}$ overlaid onto the PAH image of NGC 6810. We also define a stellar disk diameter, $D_{4.5 \mu m}$, taken as the major axis of a 0.175 MJy sr$^{-1}$ contour on the IRAC 4.5 $\mu$m image. $D_{25}$ is linearly related to $D_{4.5 \mu m}$, as shown in Figure~\ref{fig:diameters}. Taking $I_{ePAH}$ together with $D_{4.5 \mu m}$ and the total extraplanar flux, $f_{ePAH}$, we define the extraplanar PAH emission characteristic height as

\begin{equation}
\label{eq:HePAH}
H_{ePAH} = \frac{f_{ePAH}}{2 \ I_{ePAH} \ D_{4.5 \mu m}}
\end{equation}

\noindent $H_{ePAH}$ can be imagined conceptually as the height of a rectangular region with all pixel values equal to $I_{ePAH}$, width $D_{4.5 \mu m}$, and total flux $f_{ePAH}$/2. The factor of 2 arises, since $f_{ePAH}$ includes flux from above and below the disk. The measured and calculated values of $I_{ePAH}$, $D_{4.5 \mu m}$, and $H_{ePAH}$ are listed in Table~\ref{tbl:results}.

$H_{ePAH}$ behaves essentially like a measure of the extraplanar PAH emission's bulk height. Since it contains both a measure of the vertical distribution of the emission ($I_{ePAH}$) and a measure of the emission's width compared to the disk (contained within $f_{ePAH}$), $H_{ePAH}$ is most sensitive to their ratio. $D_{4.5 \mu m}$ is less important to the behaior of $H_{ePAH}$, since $f_{ePAH}$ also depends somewhat on the galaxy's diameter. We illustrate the behavior of $H_{ePAH}$ under different scenarios in Figure~\ref{fig:H_ePAH_behave}.

In order to compare $H_{ePAH}$ with the star formation properties of the host galaxy, we converted $L_{IR}$ into SFR using equation 3 from \cite{ken98}:

\begin{equation}
\label{eq:Kennicutt_98}
\frac{SFR}{1 \ M_{\odot} \ yr^{-1}} = \frac{L_{IR}}{5.8 \times 10^9 \ L_{\odot}}
\end{equation}

\noindent Then using this value along with the $D_{25}$ value (see Table~\ref{tbl:sample}), we calculated the SFR surface density:

\begin{equation}
\label{eq:Sigma_SFR}
\Sigma_{SFR} = \frac{4 \ SFR}{\pi \ D_{25} {}^2}
\end{equation}

\noindent We list the $\Sigma_{SFR}$ values in Table~\ref{tbl:results}.
 
Since $\Sigma_{SFR}$ measures an average over the entire stellar disk while star formation tends to concentrate in certain regions, a global calculation will typically underestimate SFR surface density. Therefore we developed a measurement of SFR surface density, which includes quantification of the star formation location and distribution. In doing so, we used the 8.0 $\mu$m band emission as an approximate, though imperfect, tracer of star formation rather than the 24 $\mu$m band emission, which is often used as such a tracer \citep{cal07}. The archival MIPS data for our sample suffers from lower resolution, bright (sometimes saturated) PSFs that overwhelm or significantly cover some galaxies making reliable disk diameters tricky to obtain, other detector artifacts that complicate photometry, and strong contamination from the AGN in some objects. These characteristics make the 24 $\mu$m band less useful than 8.0 $\mu$m in determining SFR surface density for our sample.

Using the same algorithm for finding $I_{ePAH}$, we found the characteristic specific intensity within the disk region of the 8.0 $\mu$m images. Although the nuclear regions of some images were compromised by the data reduction process, only small groups of pixels are affected compared to the total numbers in the disks, so they have a negligible effect on the median disk pixel value. Overlaying a contour at this value approximately outlines the densest region or regions of star formation. The contour produced alternatively a concentrated region surrounding the nucleus (e.g. NGC 6810, M82), a region spanning nearly the entire disk (e.g. NGC 4945), or several distinct regions and clumps within the disk (e.g. NGC 55, NGC 3628). Measuring the diameter of a single region or summing the diameters of multiple regions parallel to the galaxy's major axis, we obtained a characteristic diameter of PAH emission in the disk, $D_{PAH}$. We then used $D_{PAH}$ to calculate a characteristic SFR surface density ($\Sigma_{SFR}^{'}$). The values of $D_{PAH}$ and $\Sigma_{SFR}^{'}$ are listed in Table~\ref{tbl:results}, and $D_{PAH}$ is plotted versus $D_{25}$ in Figure~\ref{fig:diameters}. As a test to justify our use of $D_{PAH}$, we also compared best-estimate, non-photometric 24 $\mu$m diameters for the seven galaxies where we could measure them with each galaxy's $D_{PAH}$ value and found a one-to-one correlation to within $\sim$ 10\%.

% ----- RESULTS & DISCUSSION -----

\section{RESULTS AND DISCUSSION}
\label{results_discuss}
We find extraplanar PAH emission features in all but one galaxy. The only exception, NGC 1705, has the least favorable orientation ($i$ = 42$^\circ$) of our sample galaxies. The maximum extent ($z_{ext}$) of the features ranges from $\sim$ 0.8 kpc (NGC 1569, NGC 5253) up to $\sim$ 6.0 kpc (NGC 891, NGC 1808, and M82). The extraplanar PAH emission features have varied morphology and spatial distribution relative to their host galaxies. Figures~\ref{fig:55} -~\ref{fig:82} show comparisons of the 4.5 and 8.0 $\mu$m images for each galaxy. Appendix~\ref{append:galaxies} contains a discussion of the PAH features for each individual galaxy.

The location and morphology of extraplanar PAH features often strongly suggests where they originated. In some cases, filaments or plumes appear to extend from the galaxy's nucleus (e.g. NGC 1482, NGC 2992), while other galaxies exhibit features extending from locations spread throughout the disk (e.g. NGC 891, NGC 4631 and NGC 4945). In each of these cases, the extraplanar features extend up from the disk, but also tend to splay outward radially. This morphology suggests a radially decreasing density gradient in the ISM at the point of origin. Some evidence exists for isolated extraplanar features at locations far from the nucleus (e.g. NGC 3628), but these appear to be outliers. For some galaxies (e.g. NGC 2992 and NGC 3628), the morphology, location, and orientation of particular PAH features indicates tidal disruption rather than wind-driven origins.

The extraplanar PAH features we identify often coincide with previously observed H$\alpha$, X-ray, UV, and molecular features. Amongst these observations, H$\alpha$ appears to trace the  spatial extent and morphology of extraplanar PAH features best, though more observations of X-ray, UV, and molecular features are needed for comparison. Bright H$\alpha$ filaments and areas of diffuse H$\alpha$ emission very frequently have a PAH analog. However there doesn't seem to be an entirely consistent correlation between these two tracers, since some significant features do not match (e.g. NGC 1705 and NGC 6810). The $z_{ext}$ of PAH emission is often less than coincident H$\alpha$ or X-ray features, suggesting PAHs do not survive out to the same distances as the wind's gas components, or the sensitivity of the IRAC data is not sufficient to pick up the full extent of these features.

We also compare our measurement of extraplanar PAH flux ($f_{ePAH}$) to the total IR flux ($F_{IR}$) of our sources. Since $F_{IR}$ relates to the galaxy's SFR (excluding possible contributions from the AGN), comparing these fluxes should give some insight into what physical mechanism drives the PAHs out of the galaxy. We plot $f_{ePAH}$ vs. $F_{IR}$ in Figure~\ref{fig:plot1}. The line in Figure~\ref{fig:plot1} represents a power law fit given by:

\begin{equation}
\label{eq:fit1}
f_{ePAH} = 0.010 \ (\pm ^{0.003}_{0.002}) \ \left(\frac{F_{IR}}{10^{-9} \ erg \ s^{-1} \ cm^{-2}}\right)^{1.13 \ (\pm 0.11)} \ Jy
\end{equation}

\noindent This correlation is thus nearly consistent with linearity and has correlation coefficient $r$ = 0.89. It provides further support for a link between extraplanar emission and star formation as well as a surface density threshold on the energy or momentum injection rate for ejection of material above the disk \citep[e.g.][]{sch85,mac88,mac89,koo92a,koo92b,leh95,ros03,str04a,str04b,tul06}. Significantly, all the open-symbol points - indicating galaxies where we used a 20-$\sigma$ contour region to measure $f_{ePAH}$ - fall below this power law fit in Figure~\ref{fig:plot1}. These points should probably be considered lower limits, since part of the extraplanar PAH emission associated with these lower inclination disks may overlap with the PAH emission from the disk or be occulted by the disk itself. The points for NGC 2992 and NGC 4388 also lie below the fit. They are the only objects in our sample clearly dominated by a central AGN rather than star formation. This result suggests AGN-driven winds in the absence of robust star formation do not eject PAHs as efficiently. In these cases, the presence of an AGN may also affect the emitting grain distribution via grain destruction or ionization by the hardness of the radiation field \citep{smi07,dia10}. This tentative difference between AGN- and supernova-driven winds needs to be confirmed with a larger sample of AGN.

We compare the characteristic properties of extraplanar PAH emission to SFR surface density in Figure~\ref{fig:plot2}. The plot in the top-left panel of Figure~\ref{fig:plot2} suggests a moderate correlation between $I_{ePAH}$ and $\Sigma_{SFR}$ with significant scatter and correlation coefficient $r$ = 0.55. The open symbols on the plot tend to fall at lower $I_{ePAH}$ values than other galaxies at similar values of $\Sigma_{SFR}$, suggesting again that an inclined disk probably coincides with or occults a significant portion of the extraplanar PAH flux. The plot in the center-left panel of Figure~\ref{fig:plot2} shows scattered $H_{ePAH}$ values, but we do find that the galaxies with smaller diameters ($D_{4.5 \mu m}$ \ \textless \ 5 kpc; NGC 1569, NGC 1705, and NGC 5253) tend to fall near the lower end of the $H_{ePAH}$ range, as would be expected. NGC 4388 also has a low $H_{ePAH}$ value relative to the rest of the sample, and it's very likely that flux from tidal features increases the value for NGC 2992, so the AGN without robust star formation in our sample tend not to produce broadly extended extraplanar PAH emission features.

As listed in Table~\ref{tbl:results}, our sample galaxies span a broad range of sizes and luminosities. Since large galaxies have more of everything compared to dwarf galaxies, they not only have larger SFRs, stellar masses, and disks on average than their dwarf counterparts, but their disks and associated extraplanar regions also extend higher vertically from the disk mid-plane than in dwarf galaxies, and so they tend to have larger $H_{ePAH}$ values on average. To put all galaxies on the same footing, it is therefore natural to normalize $H_{ePAH}$ by the size of the galaxy. Normalization of $H_{ePAH}$ by $D_{25}$ results in a significant correlation with $\Sigma_{SFR}$, which is shown in the bottom-left panel of Figure~\ref{fig:plot2}. The correlation coefficient between $H_{ePAH}$/$D_{25}$ and $\Sigma_{SFR}$ is $r$ = 0.93, and the best fit power law has an exponent of 0.62 $\pm$ 0.12.

The point for NGC 891 (read vertically from the labels in the center-left panel) exhibits the largest offset from the overall trend. The anomalously large $H_{ePAH}$ for NGC 891 is more challenging to explain, but may be due to the halo-like structure of the extraplanar PAH emission in contrast with most other galaxies in our sample, which display filamentary extraplanar PAH emission. In the case of halo-like PAH emission, equation~\ref{eq:HePAH} produces a larger $H_{ePAH}$ value than in the case of filamentary emission, because the overall flux ($f_{ePAH}$) is larger and $I_{ePAH}$ is smaller (see Figure~\ref{fig:H_ePAH_behave}).

Using $D_{PAH}$ to calculate $\Sigma_{SFR}^{'}$ in the right column of Figure~\ref{fig:plot2}, we find similar results to the left column of plots. The correlation coefficient between $H_{ePAH}$/$D_{25}$ and $\Sigma_{SFR}^{'}$ is $r$ = 0.89, and the best fit power law shown in the bottom-right panel of Figure~\ref{fig:plot2} has an exponent of 0.40 $\pm$ 0.12. We also found a similar correlation between $H_{ePAH}$/$D_{4.5 \mu m}$ and $\Sigma_{SFR}$. As a note of caution to the reader, we point out that the normalization factor $D_{25}$ on the vertical axis of Figure~\ref{fig:plot2} may introduce an artificial correlation due to the linear correlations between $D_{25}$, $D_{4.5 \mu m}$, and $D_{PAH}$ (see Figure~\ref{fig:diameters}) and the fact that $\Sigma_{SFR}$ and $\Sigma_{SFR}^{'}$ depend on $D_{25}^{-2}$ and $D_{PAH}^{-2}$ respectively. While we cannot completely rule out this possibility, we note that $H_{ePAH}$ also depends on $D_{4.5 \mu m}^{-1}$ according to equation~\ref{eq:HePAH}, so we would expect a log-log slope of unity between $H_{ePAH}$/$D_{25}$ and $\Sigma_{SFR}$ or $\Sigma_{SFR}^{'}$, which is inconsistent with our findings. Therefore, we believe that the correlations between $H_{ePAH}$/$D_{25}$ and $\Sigma_{SFR}$ and $\Sigma_{SFR}^{'}$ are physically motivated. These correlations support the idea of a surface density threshold on the energy or momentum injection rate necessary for elevating the disk ISM sufficiently high above the disk to be detectable as extraplanar material.

% ----- SUMMARY -----

\section{SUMMARY}
\label{summary}
We have used existing data from the {\em Spitzer Space Telescope} archive to perform the first multi-object search for extraplanar PAH emission associated with known galactic winds. The data from the {\em Spitzer} archive are ideal for detecting PAHs due to the strength of the 7.7 $\mu$m emission feature, which falls conveniently in the IRAC 8.0 $\mu$m band. Our sample contains 16 nearby known wind galaxies, for which there are publicly available IRAC data. Our analysis of these data has yielded the following results:

\begin{itemize}

\item {\em Frequent Detection} - We found extraplanar PAH features extending from 15 of the 16 known wind galaxy in our sample with the exception of NGC 1705, which does not have a favorable orientation to search for extraplanar PAH emission.

\item {\em Kiloparsec-Scale Extent} - The extraplanar PAH features we identified extend to projected distances $\sim$ 0.8 - 6.0 kpc. However, the extent of extraplanar PAH features is often less than that of coincident ionized gas features, suggesting PAHs do not survive out to the same distances, or the {\em Spitzer} data are not sensitive enough to detect PAH emission out that far.

\item {\em Varied Morphology} - A variety of extraplanar PAH features were observed from filamentary structures and large areas of diffuse emission to the rare isolated cloud. The brightest filaments and bulk extraplanar PAH emission appear to originate close to galactic nuclei.

\item {\em Correlations with Other Phases of the ISM} - Since extraplanar PAH features often trace the wind structures detected in H$\alpha$ and X-rays, the PAHs likely represent cold ISM being swept out of the galaxy by the wind. However, not all features of the ionized gas phase (particularly H$\alpha$) match perfectly, so the presence of ionized gas in a wind does not necessarily imply the presence of extraplanar PAHs. This may be due to dust grain destruction or insufficient sensitivity of the Spitzer data.

\item {\em Connection with Star Formation Activity in the Disk} - We found that extraplanar PAH emission flux correlates linearly with total IR flux, a direct tracer of the SFR (excluding possible contributions from an AGN). Employing a more quantitative analysis of the extent of extraplanar PAH emission, we derive a characteristic scale height of the extraplanar PAH emission, $H_{ePAH}$, for each galaxy. We find a significant correlation between the surface density of star formation in the disk and the characteristic scale height of the extraplanar PAH emission, once it is normalized to the galaxy diameter. This direct link between the star formation activity in the disk and the extent of the extraplanar PAH emission is similar to that seen when considering the extraplanar H$\alpha$ emission. This result reinforces the idea of a surface density threshold on the energy or momentum injection rate needed to eject material sufficiently high above the disk to be detectable as extraplanar material. 
\end{itemize}

% ----- ACKNOWLEDGEMENTS -----

\section*{ACKNOWLEDGEMENTS}
This work is based in part on observations made with the {\em Spitzer Space Telescope}, which is operated by the Jet Propulsion Laboratory, California Institute of Technology under a contract with NASA. Support for this work was provided by NASA through {\em Spitzer} archival grant 1310149. SV also acknowledges partial support from a Senior NPP Award held at the NASA Goddard Space Flight Center and from the Humbolt Foundation to provide funds for a long-term visit at MPE in 2012. We acknowledge the helpful comments of the referee. We thank Massimo Ricotti and Stacy McGaugh for comments on an early version of the manuscript, Michael McDonald for his suggestions regarding our electronic banding removal algorithm and his help with "asinh" scaled images, Alberto Bolatto for his advice regarding scale height determination, and Jonathan Fraine for his help with coding in PyRAF. We also thank G. Bicknell, J. Bland-Hawthorn, J. Cooper, C. Engelbracht, M. Regan, and R. Sutherland, who were co-Is on the original {\em Spitzer} archival proposal.

% ----- INDIVIDUAL GALAXY DISCUSSION -----
\newpage
\appendix
\section{NOTES ON INDIVIDUAL GALAXIES}
\label{append:galaxies}

\begin{itemize}
\item NGC 55 (Figure~\ref{fig:55}): Faint PAH filaments extend from the southern edge of the disk's nuclear region in the 8.0 $\mu$m image, bracketing the eastern and western extent of the nuclear region. Some evidence also exists to suggest the base of a north eastern loop or shell structure similar in location to extraplanar ones observed in [O III] by \cite{gra82} and H$\alpha$ by \cite{hoo96} and \cite{ott99}. In contrast with the H$\alpha$ observations of \cite{fer96} and \cite{hoo96}, we do not observe significant extraplanar PAH emission north of galactic disk.

\item NGC 253 (Figure~\ref{fig:253_891}, top): Complexes of filamentary and halo-like PAH structures extend both north and south from the nuclear region of NGC 253. Two particularly prominent filaments extend north and south-east from the north-eastern area of the disk. The filament extending north appears to have an 'S'-like shape with two foreground stars superimposed over the upper portion of the 'S'. The PAH emission from the filament to the south-east extends away from the nuclear region $\sim$5 kpc, making and angle of $\sim$65$^{\circ}$ with the outer part of the disk to the north-east. This same feature appears in H$\alpha$ and X-ray \citep{str02} as well as UV images \citep{hoo05}. The IRAC 8.0 $\mu$m image from this work exhibits more extraplanar emission south of the disk, whereas previous observations at other wavelengths \citep{pie00,str02,str04a,str04b,hoo05} show more prominent emission to the north. Some faint, residual banding which could not be removed is present in the 8.0 $\mu$m image but does not contribute significantly to the total extraplanar flux.

\item NGC 891 (Figure~\ref{fig:253_891}, bottom): A nearly disk-wide halo of extraplanar PAH emission extends approximately equal distances ($\sim$5 kpc) to the east and west of NGC 891's disk. The disk-wide extent of the extraplanar PAH emission matches the morphology of extraplanar dust filaments observed by \cite{how97} near the disk in optical images and H$\alpha$ and [N {\sc ii}] emission observed by \cite{mil03}. In contrast, X-ray maps exhibit an asymmetry to the extraplanar emission preferentially towards the northwest with a $\sim$6 kpc filament extending from the nuclear region in that direction \citep{bre94,bre97,str04a}. Detailed inspection of the halo region reveals finger-like PAH features suggesting the presence of Rayleigh-Taylor instabilities, and reminiscent of the 'highly structured dust features' observed by \cite{how97,how00}. Near the south-east edge of the disk, a rather prominent filament appears to be encircled by a faint loop of PAH emission. Faint streaks parallel to the western edge of the disk and separated by about 4 and 5 kpc from the galaxy are due to ghost images of the galaxy's disk in the data, but these artifacts do not contribute significantly to the total extraplanar flux. As with NGC 253, some faint residual banding remains, but also does not contribute significantly to the total extraplanar flux.

\item NGC 1482 (Figure~\ref{fig:1482_1569}, top): There are several filamentary PAH features extending from both sides of the disk in the 8.0 $\mu$m image. These features seem to extend approximately radially from regions near the nucleus. Previous H$\alpha$ observations \citep{ham99,vei02} found an 'hourglass' morphology, consistent with a biconical outflow. The X-ray observations of \cite{str04a,str04b} showed a complementary, similar morphology. While there are some indications of similar structure in the 8.0 $\mu$m IRAC image, the PAH emission features seem more evenly distributed with filaments splayed out in all directions. This complementary, but contrasting morphology between H$\alpha$ and PAH emission recalls the similar observations of M82 by \cite{eng06}. The data reduction for NGC 1482 was particularly challenging, since it's bright nucleus is less like a point source than others in our sample. The large solid angle of high luminosity within the nuclear region leads to a smearing of the PSF wings and therefore more overlap with extraplanar regions. Our fit to the wings caused by the bright nucleus took into account the smearing. Therefore we were able to remove the vast majority of the PSF wings, but the black crosshairs artifact in Figure~\ref{fig:1482_1569} resulted from slight over-subtraction along the vertical and horizontal axes of the detector.

\item NGC 1569 (Figure~\ref{fig:1482_1569}, bottom): This galaxy exhibits PAH features which, notably, do not all extending from the nuclear region or perpendicular to the disk region's major axis. Several of the brightest, large scale features extend from regions near the edges of the disk. In particular, a filament extends southwest from the western edge of the galaxy (the 'arm' identified by \cite{wal91}), and two filaments from the eastern edge of the galaxy curl towards the northeast in the image. Much of the extraplanar PAH emission morphology matches H$\alpha$ features previously observed by \cite{hod74,wal91,hun93}, though some differences can also be observed - primarily the smaller extent of the extraplanar PAH features, and their more uniformly distributed emission as compared with the more filamentary H$\alpha$ features. The bright star to the north of the galaxy generated particularly bright PSF wings, so it's removal in the 4.5 $\mu$m image left some artifacts, such as the half-moon shaped feature just north of the star. Also, some very faint pixels north of the galaxy in the 8.0 $\mu$m image coincide with a region where banding could be expected even after our best efforts at correction (near the top edge of the 8.0 $\mu$m image: Figure~\ref{fig:1482_1569}, bottom right), so the authenticity of these as PAH emission is questionable at best.

\item NGC 1705 (Figure~\ref{fig:1705_1808}, top): We did not find significant extraplanar PAH features in comparing the IRAC images. This result is somewhat surprising considering the strength of the bipolar H$\alpha$ emission observed by \cite{meu89,meu92}. The 8.0 $\mu$m image exhibits clumpy star forming regions similar to those observed by \cite{mel85}.

\item NGC 1808 (Figure~\ref{fig:1705_1808}, bottom): Due to the lower inclination of NGC 1808, the portions of extraplanar PAH features closest to the disk coincide with or are occulted by the disk, making them more difficult to distinguish. The lower inclination also causes an underestimate in the extraplanar PAH flux measurement. Extraplanar PAH features extend both from the northeastern and southwestern edges of the disk. Several filaments can be observed within a more diffuse halo of PAH emission which appears more prominent on the northeastern side of the disk. The brightest of these filaments appear similar to optical absorption features observed by \cite{phi93}, and the diffuse PAH halo is somewhat similar to the H$\alpha$ morphology described by \cite{sha10}. The tidal tails to the northwest and southeast also exhibit bright PAH emission. Some residual banding remains both to the east and west of the disk (see masking in Figure~\ref{fig:flux}).

\item NGC 2992 (Figure~\ref{fig:2992_3079}, top): In tracing the PAH emission, we recovered structures similar to the biconical outflow identified in H$\alpha$ by \cite{vei01}, but unlike radio \citep{war80,hum83,col96,gal06} and X-ray \citep{col98} maps, which primarily show extended features east of the nucleus. Particularly, filamentary PAH emission features extend out from the nucleus in a nearly radial direction (most easily seen as dark blue features in Figure~\ref{fig:2992_3079}). The northern tidal tail due to NGC 2992's interaction with NGC 2993 (not shown in the field of view) does coincide with the extraplanar PAH features on the west side of the disk, but the different morphologies allow for the distinction of filaments versus tidal tail. Emission from the coincident tidal tail to the west as well as the tidal filament extending almost directly south contributes to the extraplanar PAH flux measurement, but these components are not easily disentangled. As with NGC 1482, we fit the PSF wings of the bright nucleus, but the image was slightly over-subtracted in certain areas (the black lines emerging from the nucleus along the minor axis).

\item NGC 3079 (Figure~\ref{fig:2992_3079}, bottom): Extraplanar PAH features extend both from the eastern and western edges of the disk. Several filaments can be observed within a bright halo of PAH emission, which is thickest out to $\sim$ 8 kpc radially from the nucleus. In particular, two PAH filaments separated by about 5 kpc along the eastern edge of the disk and bracketing the nuclear region (light blue in Figure~\ref{fig:2992_3079}) as well as another filament south of the nucleus along the western edge of the disk extend outward similar to the X-shaped morphology observed by \cite{hec90} in H$\alpha$ and by \cite{dah98,pie98,str04a} in X-ray maps. We do not observe any PAH features near the nuclear region suggestive of the well-known superbubble which resides there \citep{for86,cec01}.

\item NGC 3628 (Figure~\ref{fig:3628}): Two prominent artifacts remain in the 8.0 $\mu$m image after the data reduction. Faint streaks parallel to the northern edge of the disk and separated by $\sim$ 2,3 and 7 kpc from the nuclear region of the galaxy are due to ghost images of the galaxy's disk in the data. These artifacts were masked and excluded from the extraplanar region during the extraplanar PAH flux measurement (see Figure~\ref{fig:flux}). Some residual banding extends north of the disk from the bright nucleus and was also excluded from the flux measurement. In addition to these artifacts in the 8.0 $\mu$m image, the filament rising at an acute angle from the northwestern edge of disk appears to extend in the same direction as the tidal tail observed by \cite{kor74}, \cite{hay79} and \cite{chr98}, suggesting it does not result from a wind. The most prominent non-artifact, non-tidal, extraplanar PAH features are a faint cloud of emission separated from the northeastern edge of the disk by about 1 kpc, a brighter protrusion from the southeastern edge of the disk and a bit closer to the nucleus, and perhaps a 2 kpc loop of PAH emission from the far southwestern edge of the disk. There may also be some faint emission coincident with the region just southwest of the nucleus where extended H$\alpha$ \citep{fab90} and outflowing molecular material \citep{irw96} have been observed, but certainly not on the same scale or bright enough for confirmation in the IRAC data. The morphology of these faint features does not match the X-ray emission seen extending from the disk in previous observations \citep{fab90,dah96,str04a}.

\item NGC 4388 (Figure~\ref{fig:4388_4945}, top): Apart from a halo which extends north and south of the disk, very little extraplanar PAH emission can be observed. Some faint residual banding similar to that of NGC 891 can be seen in the 8.0 $\mu$m image, but it does not contribute significantly to the total extraplanar flux. Also, the fainter band in the disk directly north of the nucleus resulted from PSF subtraction, and not any real feature in the data. We see little to no evidence of extraplanar PAH emission coincident with the region where \cite{vei99} observed a complex of highly ionized gas extending north of the nucleus. Nor do we see a PAH analog to the extended, nuclear radio emission \citep{hum83,sto88}.

\item NGC 4631 (Figure~\ref{fig:4631}): Two $\sim$ 3-4 kpc-scale filaments extend from the southeastern edge of the disk. Some fainter protrusions show up just south of the nuclear region. Also, a dense region of apparent Rayleigh-Taylor instability 'fingers' and filaments covers the entire central portion of the northern disk edge. The PAH filaments north of the disk appear brighter than any features south of the disk, which matches well with previous H$\alpha$ and X-ray observations \citep{dah98,wan01,str04a}.

\item NGC 4945 (Figure~\ref{fig:4388_4945}, bottom): Filamentary extraplanar PAH features extend from all regions of this galaxy's disk. The southeastern edge of the disk exhibits the highest concentration of filamentary features, where a dense region of plumes is bright against the background. Filaments northwest and directly north of the nuclear region extend up from the disk and radially outward suggesting similar structure to the wide cone-like feature observed in broadband optical emission by \cite{nak89} and in line splitting by \cite{hec90} (perhaps better seen in the optical image shown by \cite{elm97}). We do not see evidence in the PAH emission for the narrower conical wind-blown cavity observed in H$\alpha$ and H {\sc ii} \citep{moo96}, as well as X-ray \citep{str04a}. The clear presence and morphology of extraplanar PAH emission filaments at large galactic radii suggests much of the extraplanar PAHs were ejected from regions well outside the starburst disk or ring traced by Br$\gamma$ and L band imaging \citep{moo96}. Since NGC 4945 is quite near the Galactic plane ($b$ $\approx$ 13$^\circ$), these PAH filaments at large galactic radii may not appear to have analogs in previous observations at other wavelengths due to Galactic absorption as noted by \cite{col98} and \cite{str04a}.

\item NGC 5253 (Figure~\ref{fig:5253_6810}, top): As with our data reduction for NGC 1482, the 8.0 $\mu$m PSF was slightly over-subtracted along the vertical and horizontal axes after fitting and subtracting the PSF wings. Also, ghost image artifacts near the nucleus appear as multiple point sources in the 8.0 $\mu$m image. Extraplanar PAH filaments seem to extend approximately radially from the nuclear region in most directions. Filaments to the east and west match up somewhat with previous observations of H$\alpha$ filaments \citep{marl95,mar95}. However, more prominent PAH emission to the southeast of the nucleus coincides with an H {\sc i} plume which has been interpreted as a tidal remnant or inflow \citep{kob08}. The filaments of PAH emission to the west of the nucleus appear to extend into the region where a "quiescent" (neither outflowing nor inflowing) H$\alpha$ bubble has been observed \citep{mar98}.

\item NGC 6810 (Figure~\ref{fig:5253_6810}, bottom): At first look, some of the radially-oriented PAH features (light blue in Figure~\ref{fig:5253_6810}) around the nucleus in the 8.0 $\mu$m image appeared conspicuously similar to the wings of the SSC-provided PRF. However, upon fitting a PSF to the galaxy's nucleus, we found that there was little to no contribution to the extraplanar features from the PSF. Therefore, the extended extraplanar PAH emission features appear as two galactic-scale bubbles emerging from either sides of NGC 6810's nuclear region. This dust morphology resembles the X-ray features observed by \cite{str07} more closely than H$\alpha$ maps, which include a prominent filament extending toward the north-west, confirmed by \cite{str07} after noting its presence in the observations of \cite{ham99}.

\item M82 (Figure~\ref{fig:82}): We recovered the same widespread extraplanar PAH emission observed by \cite{eng06} and also found a consistent value for the extent of the features ($z_{ext}$ = 6.0 kpc). As discussed by \cite{eng06}, the extraplanar PAH emission is not confined to the biconical wind region as defined by the extraplanar H$\alpha$ \citep[e.g.,][]{leh99}, X-ray \citep{str04a}, and UV \citep{hoo05} emission, but within that region it has similar structure to the tracers at other wavelengths. See the discussion of \cite{eng06} for more detail.

\end{itemize}

\newpage

% ------ REFERENCES ------

\newpage
\normalsize

% ----- FIGURES -----

% ROSSA & DETTMAR ANALOG PLOT

% ELECTRONIC BANDING

% PSF FITTING AND SUBTRACTION

% STELLAR CONTINUUM SUBTRACTION

% SCALE HEIGHT EXPONENTIAL FITTING

% SCALE HEIGHT DISK REGION

\FloatBarrier

% FLUX CALCULATION REGIONS
\begin{figure}[htbp]
\centering
{\includegraphics[width=1.0\textwidth]{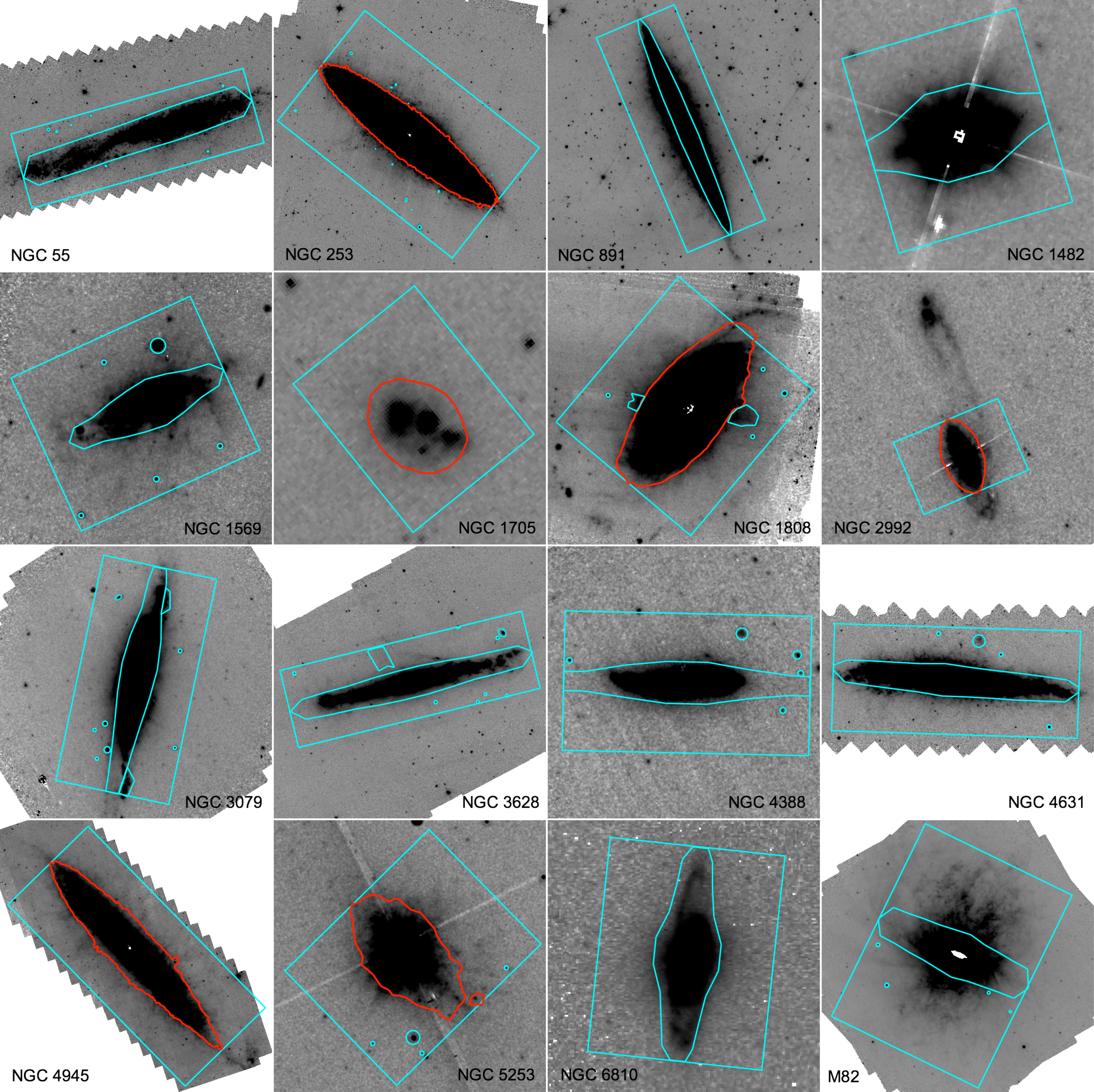}}
\caption{\footnotesize{Regions used to determine the flux of the extraplanar PAH emission: IRAC 8.0 $\mu$m maps overlaid with the disk, 20-$\sigma$, mask, and extraplanar regions as described in \S~\ref{flux}. 20-$\sigma$ contours are shown in red. Note that 20-$\sigma$ and disk regions were generated from the IRAC 4.5 $\mu$m images, so the 8.0 $\mu$m disk emission will often differ significantly in spatial extent (e.g. M82). The intensity scalings are logarithmic. North is up and east is left in all images.}}
\label{fig:flux}
\end{figure}

% H_ePAH CALCULATION
\begin{figure}[htbp]
\centering
\includegraphics[width=0.7\textwidth]{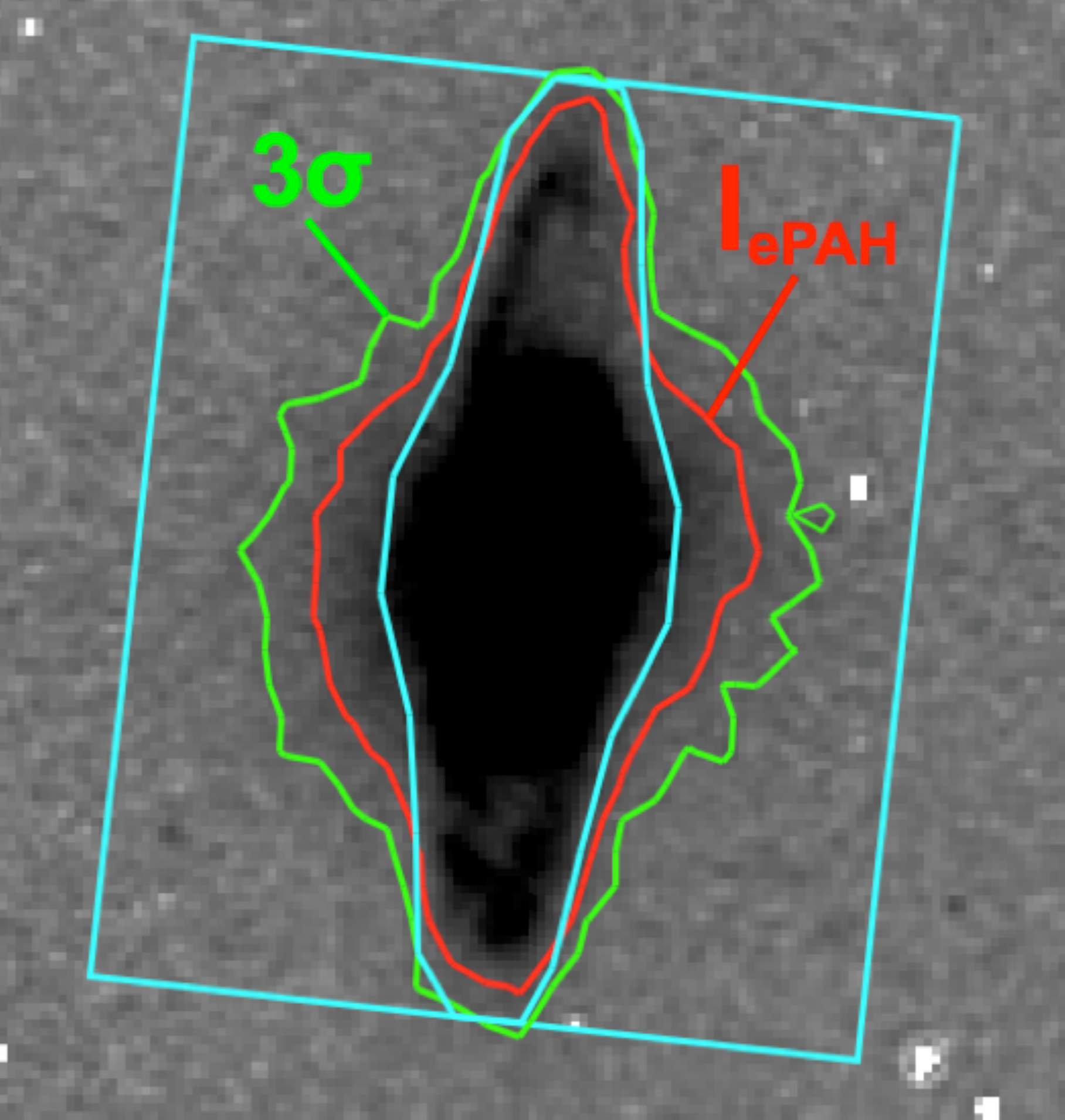}
\caption{\footnotesize{Determining the characteristic extraplanar emission height, $H_{ePAH}$. The overlaid map is the PAH image of NGC 6810. The green contour is 3-$\sigma$ above the background, and the red contour follows the characteristic intensity value, $I_{ePAH}$. The cyan regions are the disk and extraplanar regions. The intensity scaling of the overlaid image is logarithmic. North is up and east is left.}}
\label{fig:H_ePAH_calc}
\end{figure}

% DIAMETER COMPARISON PLOT
\begin{figure}[htbp]
\centering
\includegraphics[width=0.65\textwidth]{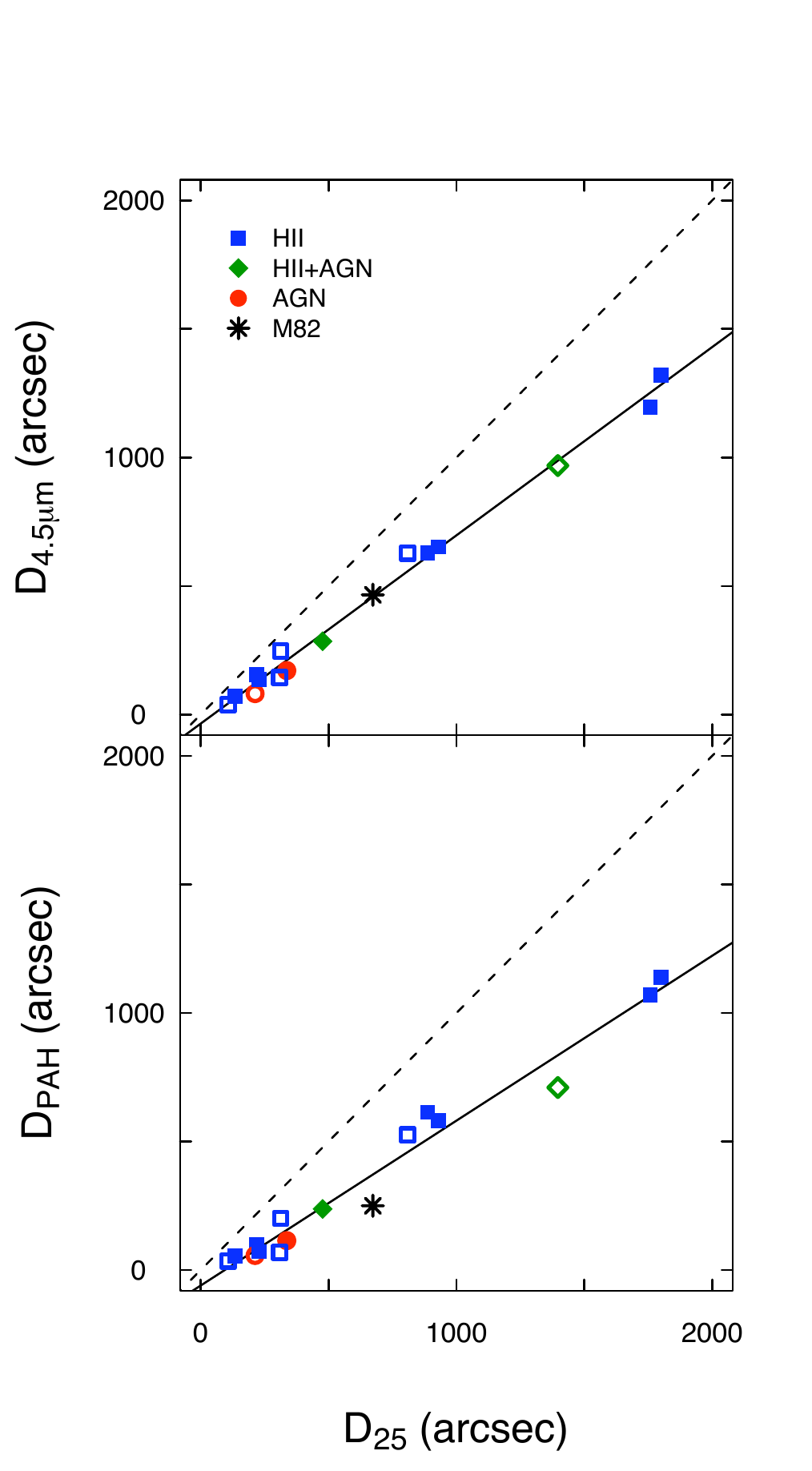}
\caption{\footnotesize{Comparisons of the mid-infrared and optical diameters: the major axis in the IRAC 4.5 $\mu$m images at the 0.175 MJy sr$^{-1}$ contour ($D_{4.5 \mu m}$) and characteristic PAH diameter ($D_{PAH}$, see \S~\ref{results_discuss}) versus major axis in B-band images at apparent magnitude $m$ = 25 ($D_{25}$). Meaning of the symbols is the same as in Figure~\ref{fig:RossaDettmar_analog}. The solid lines in the top and bottom panels are linear fits with slopes $\alpha$ = 0.73 and $\alpha$ = 0.64, respectively. The dashed lines shows the 1-to-1 relationship for reference.}}
\label{fig:diameters}
\end{figure}

% H_ePAH BEHAVIOR
\begin{figure}[htbp]
\centering
\includegraphics[width=0.8\textwidth]{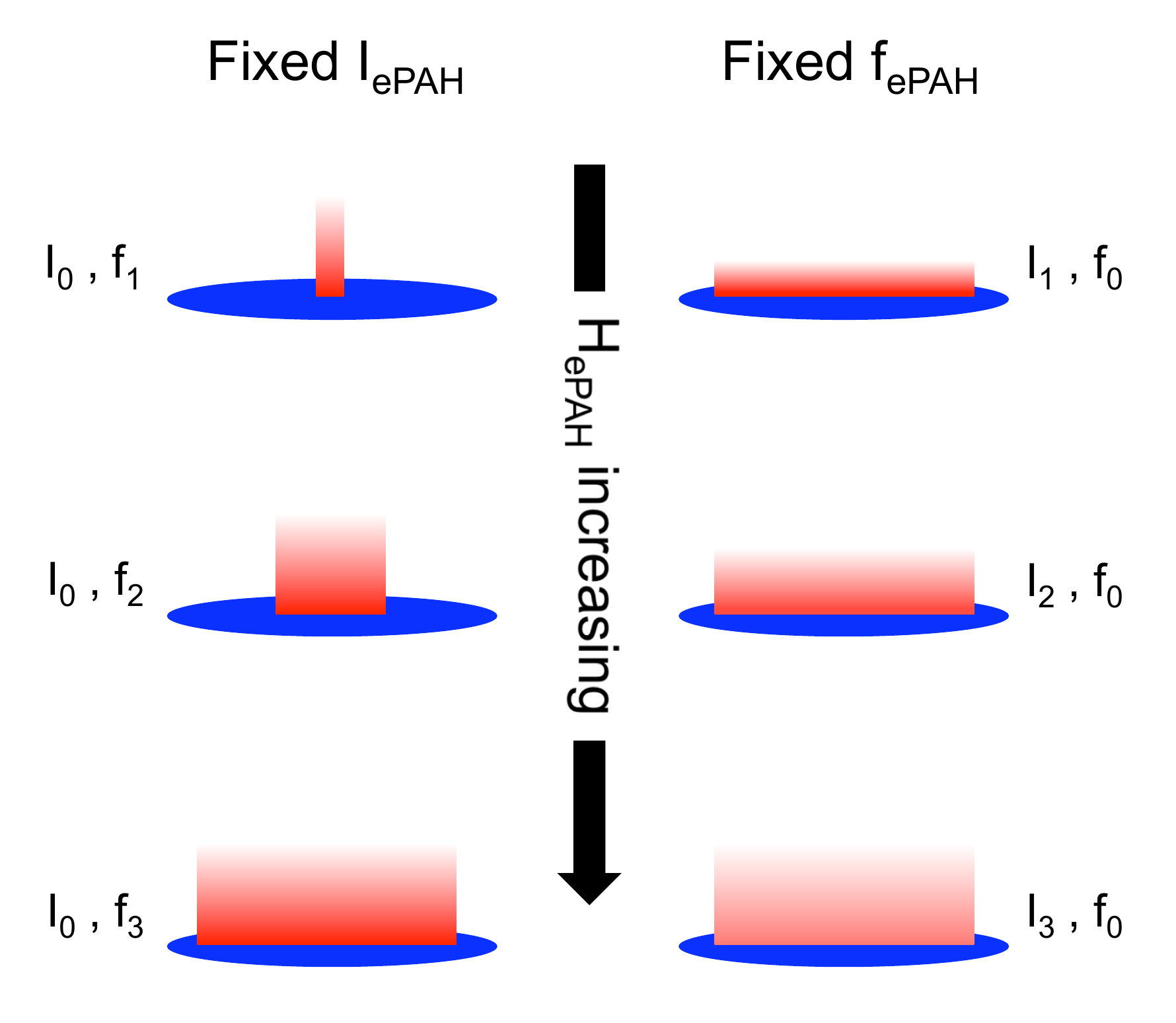}
\caption{\footnotesize{The behavior of the characteristic extraplanar PAH emission height, $H_{ePAH}$, under different scenarios. Blue ellipses represent the stellar disk of the galaxy, and the rectangular red gradients represent the extraplanar emission. Left: for fixed values of the characteristic extraplanar intensity, $I_{ePAH}$, larger values of total extraplanar PAH flux, $f_{ePAH}$ ($f_1$  \textless \ $f_2$ \textless \ $f_3$) produce larger values of $H_{ePAH}$ (same vertical profile, but more width). Right: for fixed values of $f_{ePAH}$, smaller values of $I_{ePAH}$ ($I_1$  \textgreater \ $I_2$ \textgreater \ $I_3$) produce larger values of $H_{ePAH}$ (same width and total flux, but graduated vertical profile).}}
\label{fig:H_ePAH_behave}
\end{figure}

\FloatBarrier

% REDUCED DATA IMAGES...

\renewcommand{\thefigure}{\arabic{figure}\alph{subfigure}}
\setcounter{subfigure}{1}

\begin{figure}[htbp]
\centering
\includegraphics[width=1.0\textwidth]{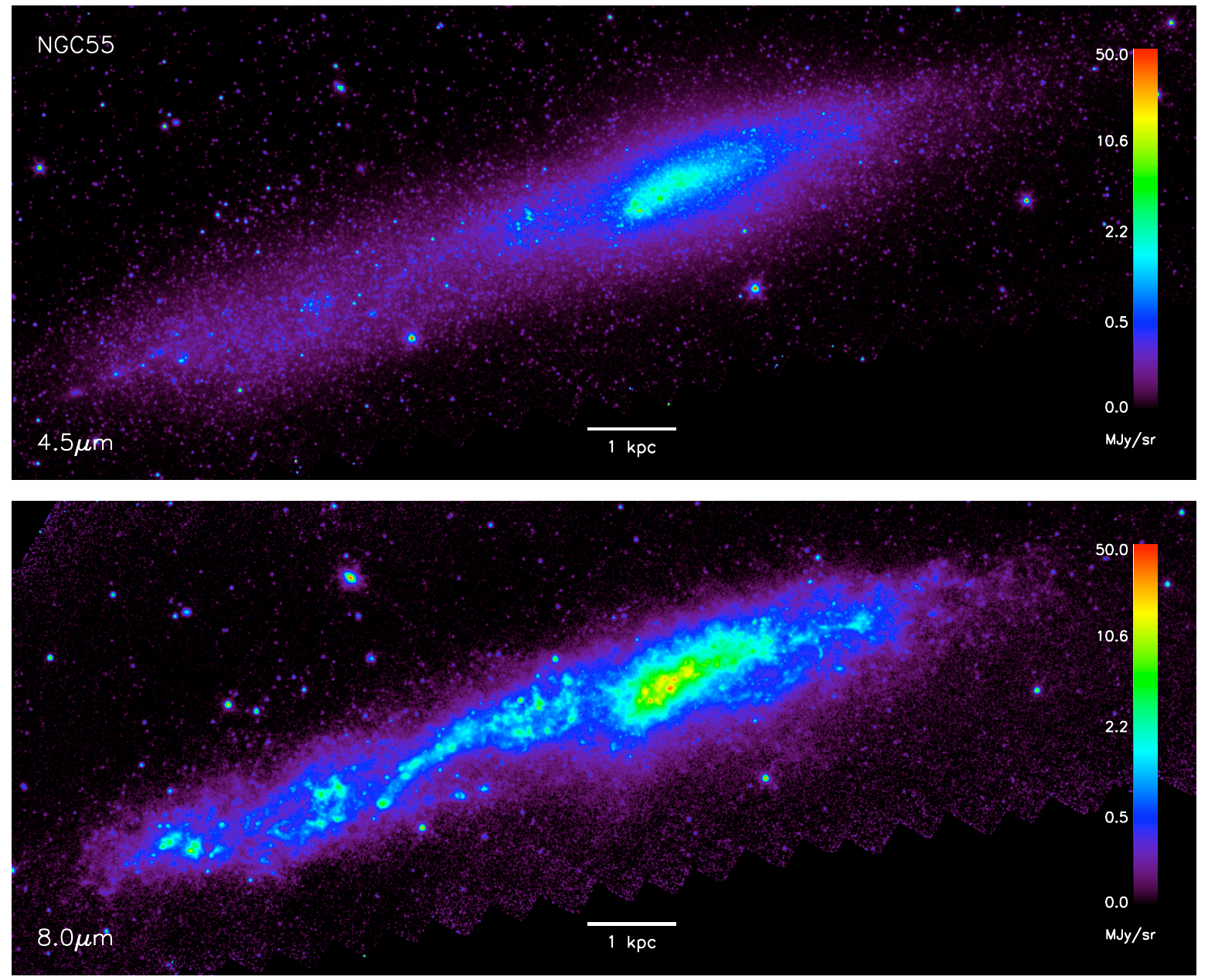}
\caption{\footnotesize{Comparing IRAC 4.5 and 8.0 $\mu$m maps of NGC 55. The images are displayed with the "asinh" (inverse hyperbolic sine) scaling from \cite{lup99}, which has approximately linear scale in the low flux range and approximately logarithmic scale in the high flux range. This scaling allows faint features to stand out without obscuring structure in the brighter regions. North is up and east is left in both images.}}
\label{fig:55}
\end{figure}

\addtocounter{figure}{-1}
\addtocounter{subfigure}{1}
\begin{figure}[htbp]
\centering
\includegraphics[width=1.0\textwidth]{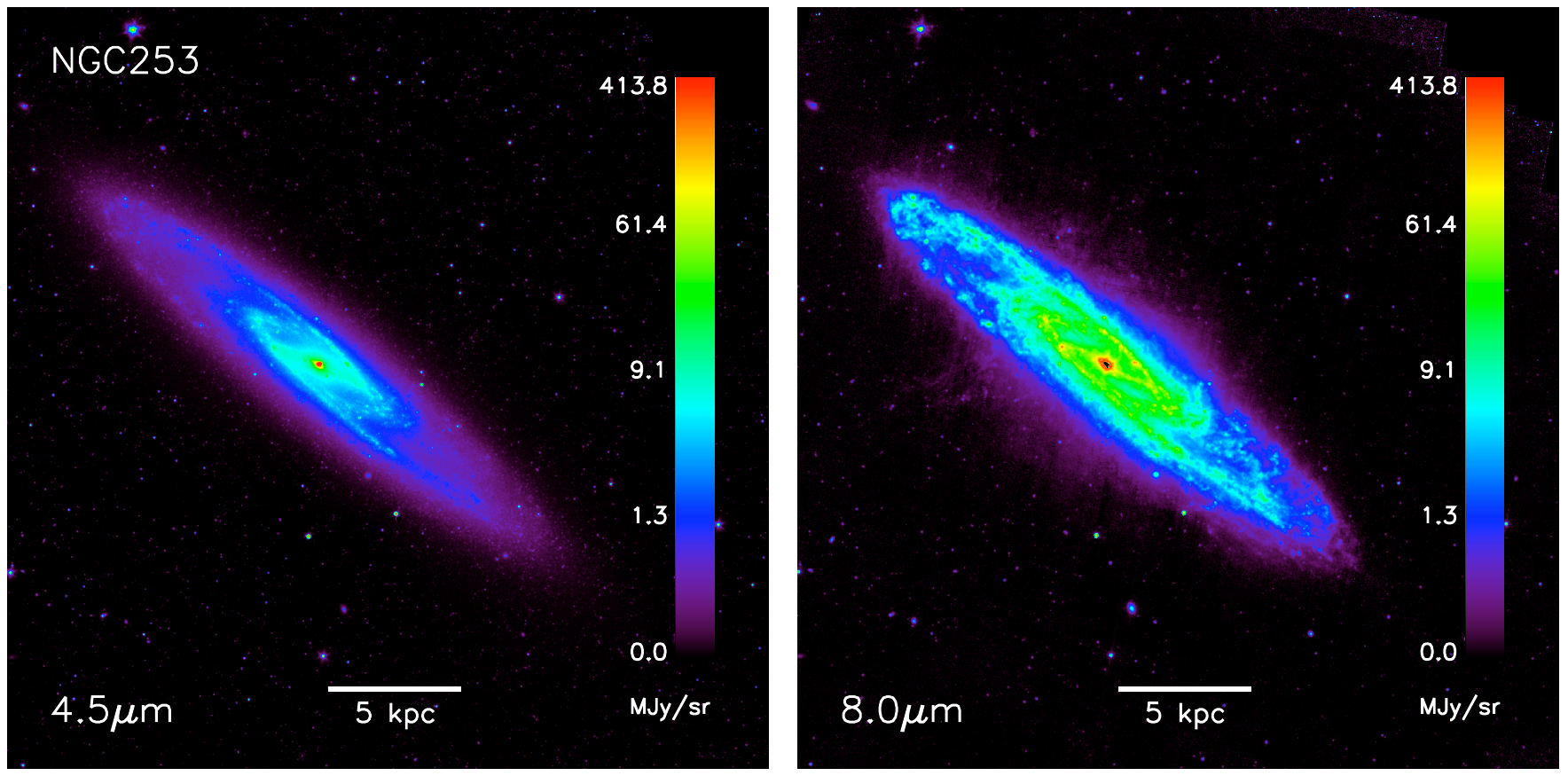}
\includegraphics[width=1.0\textwidth]{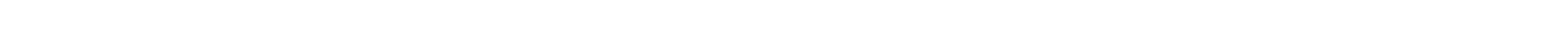}
\includegraphics[width=1.0\textwidth]{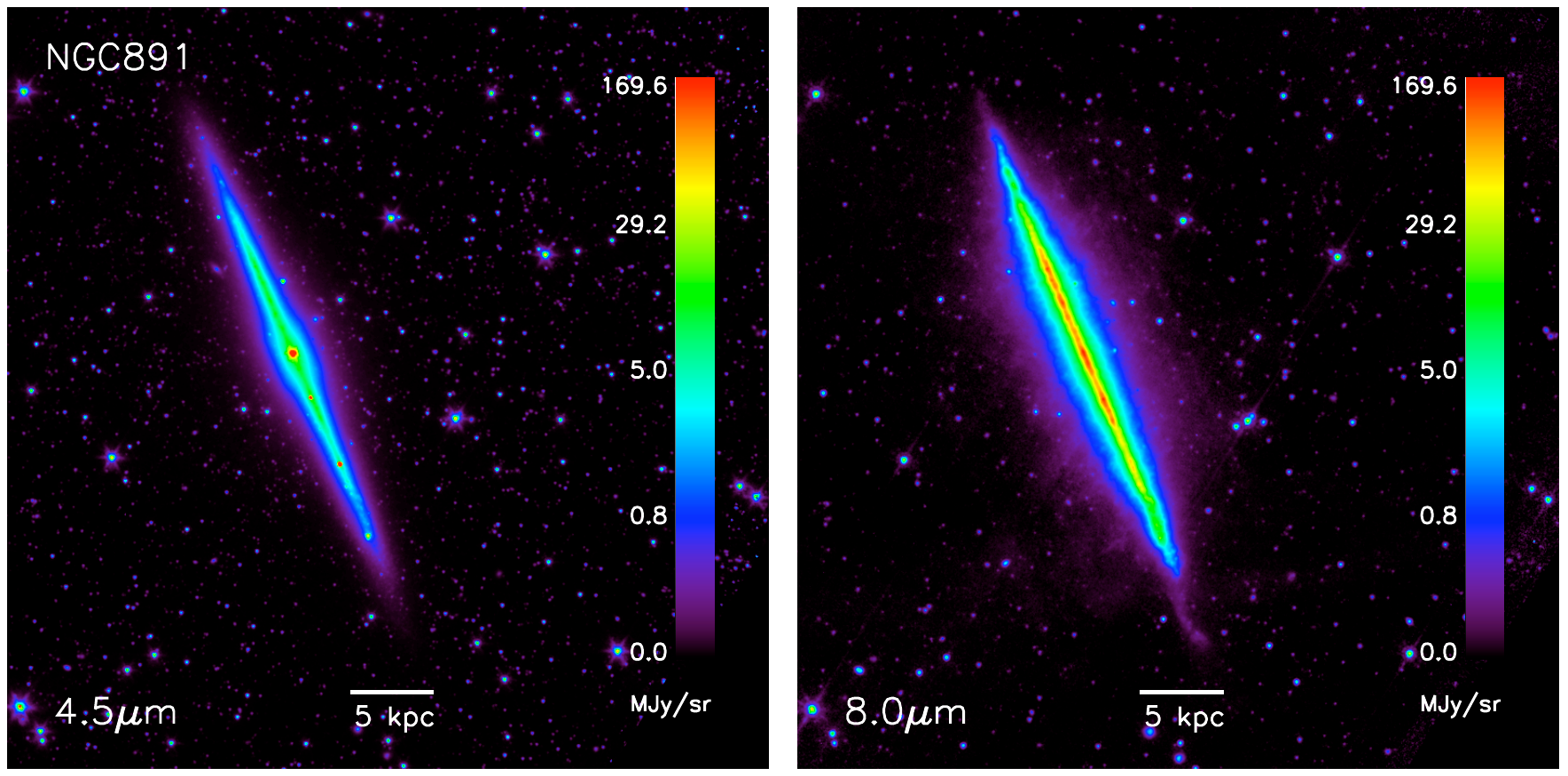}
\caption{\footnotesize{Comparing IRAC 4.5 and 8.0 $\mu$m maps of NGC 253 and NGC 891. The intensity scalings are "asinh" as described in Figure~\ref{fig:55}. North is up and east is left in all images.}}
\label{fig:253_891}
\end{figure}

\addtocounter{figure}{-1}
\addtocounter{subfigure}{1}
\begin{figure}[htbp]
\centering
\includegraphics[width=1.0\textwidth]{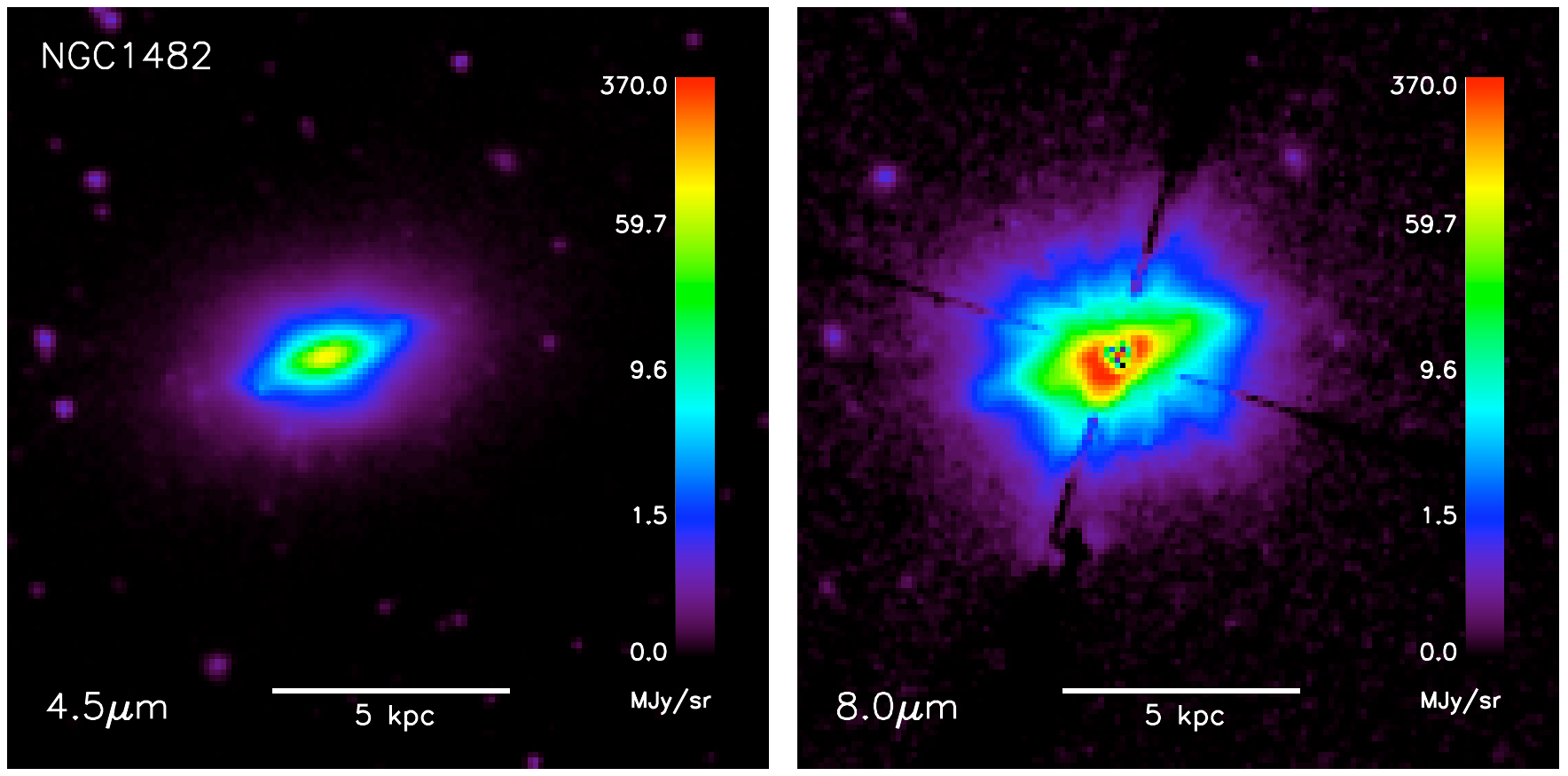}
\includegraphics[width=1.0\textwidth]{spacer}
\includegraphics[width=1.0\textwidth]{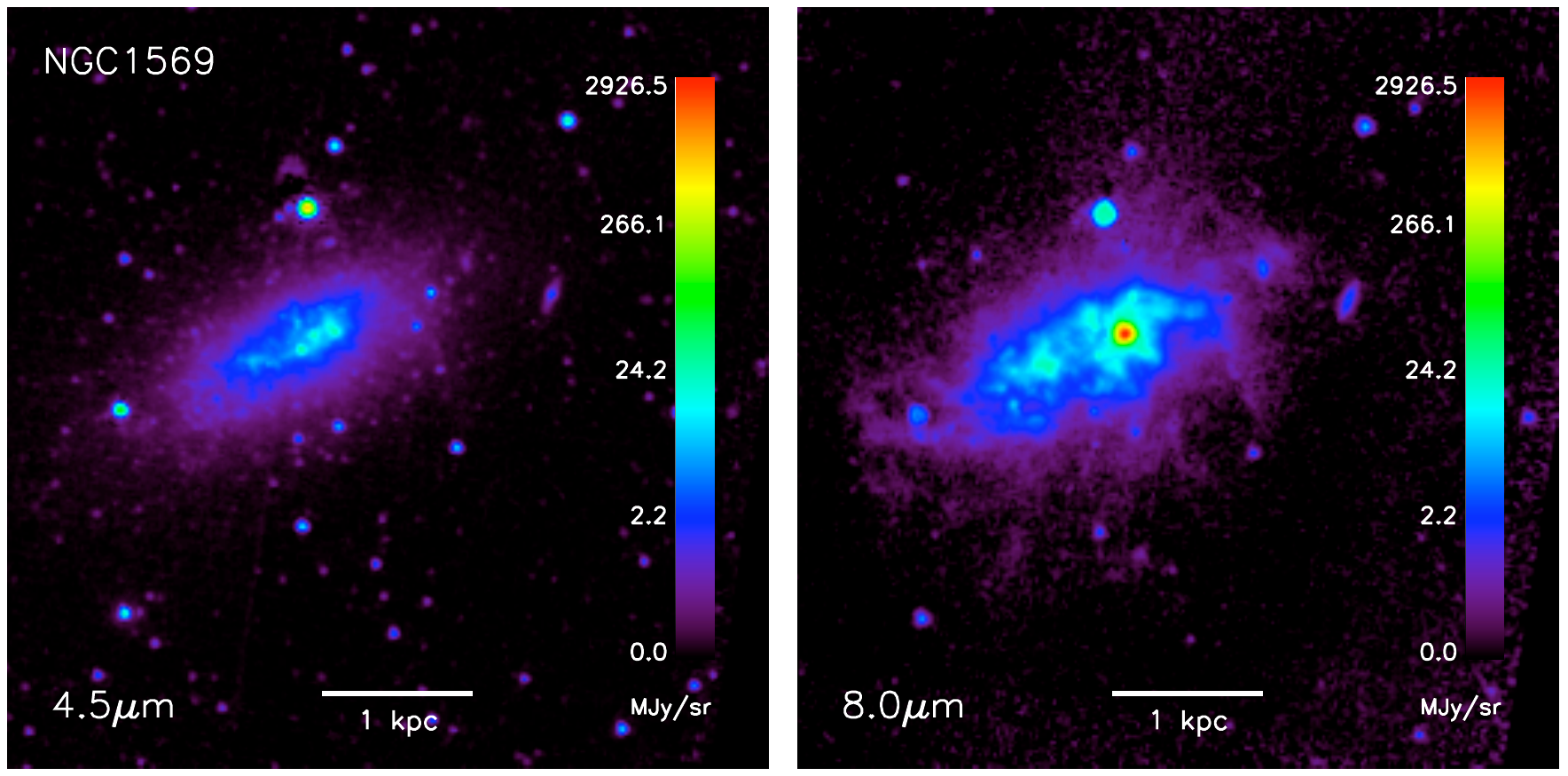}
\caption{\footnotesize{Comparing IRAC 4.5 and 8.0 $\mu$m maps of NGC 1482 and NGC 1569. The intensity scalings are "asinh" as described in Figure~\ref{fig:55}. North is up and east is left in all images.}}
\label{fig:1482_1569}
\end{figure}

\addtocounter{figure}{-1}
\addtocounter{subfigure}{1}
\begin{figure}[htbp]
\centering
\includegraphics[width=1.0\textwidth]{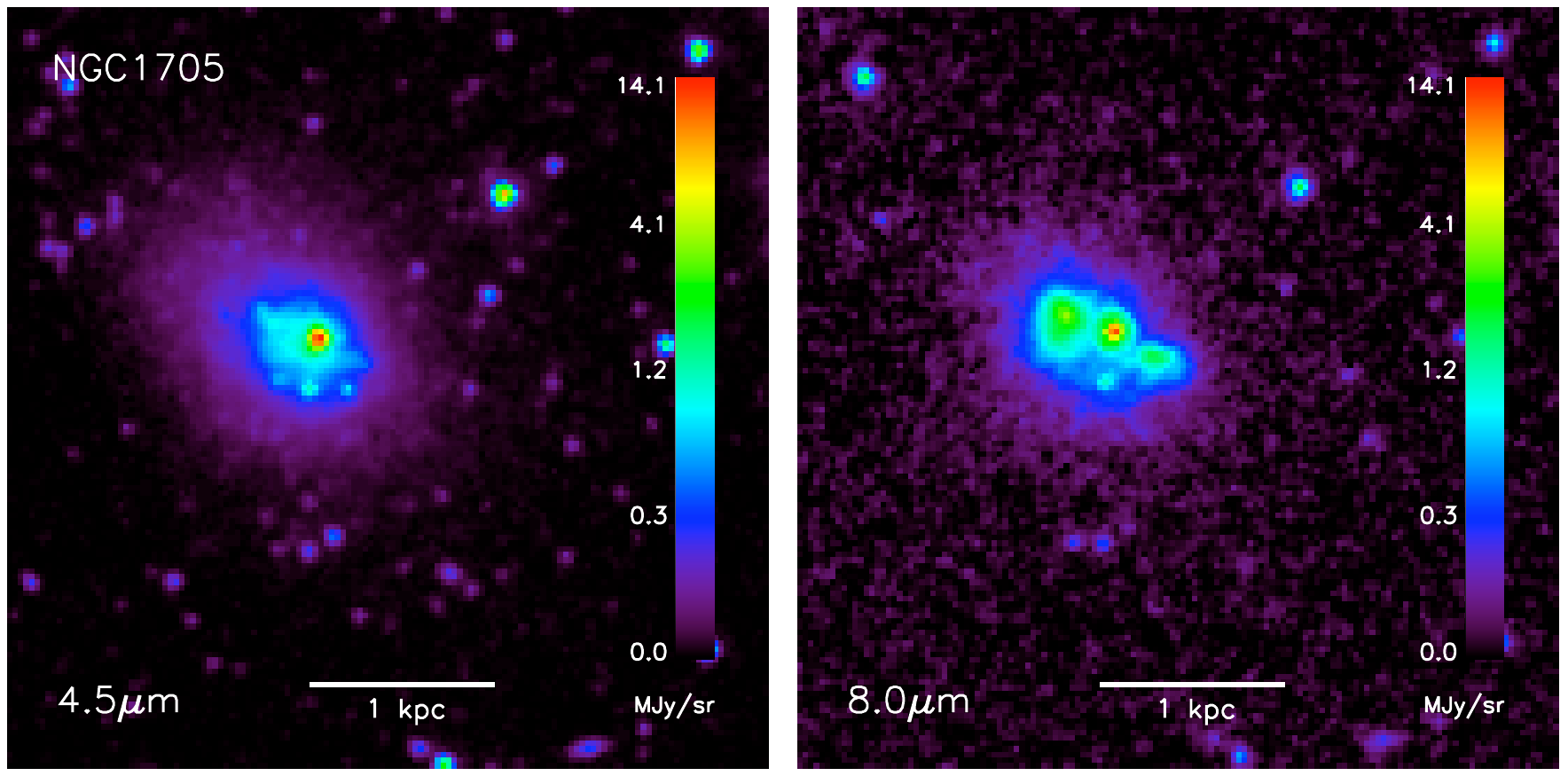}
\includegraphics[width=1.0\textwidth]{spacer}
\includegraphics[width=1.0\textwidth]{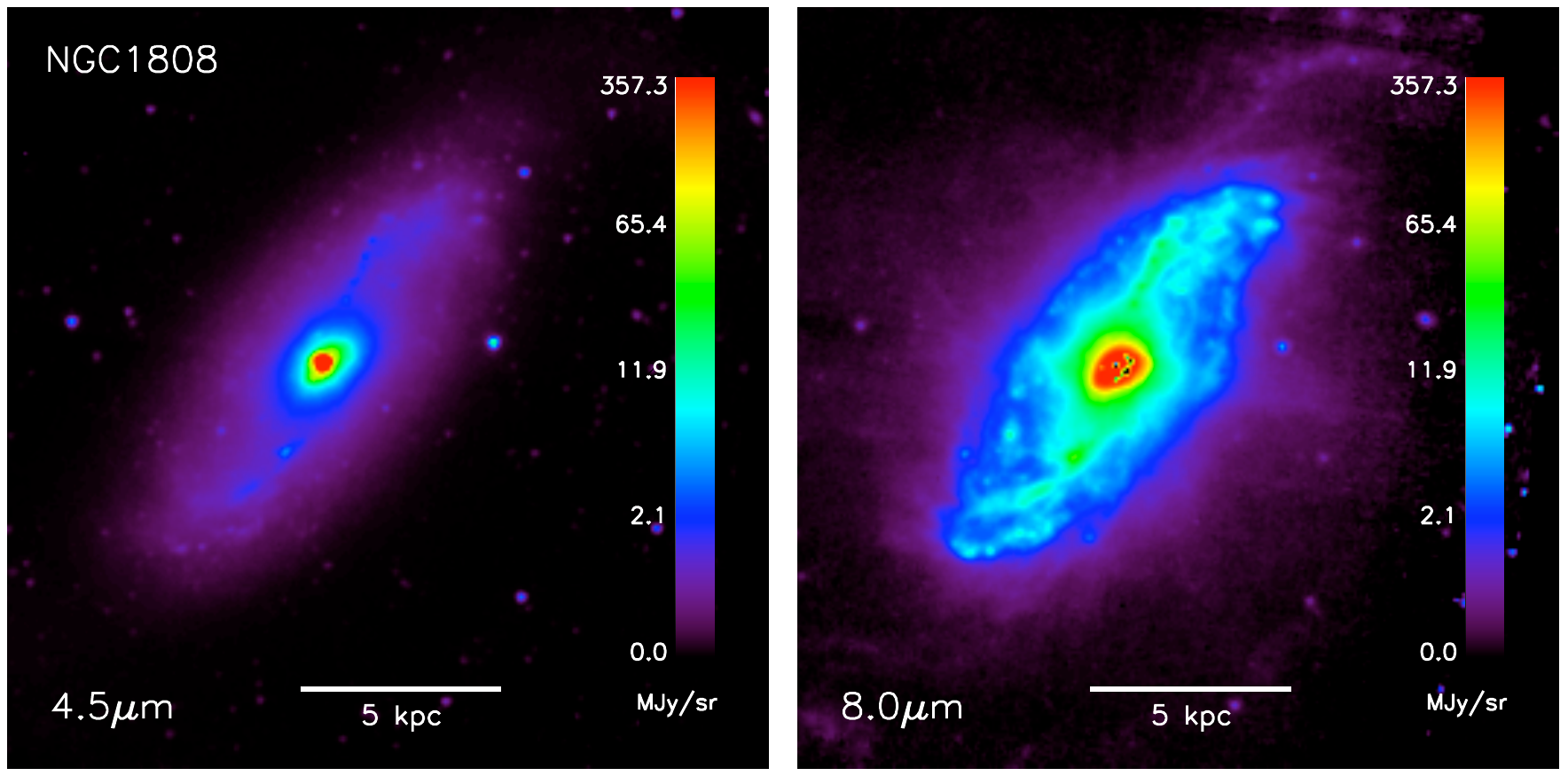}
\caption{\footnotesize{Comparing IRAC 4.5 and 8.0 $\mu$m maps of NGC 1705 and NGC 1808. The intensity scalings are "asinh" as described in Figure~\ref{fig:55}. North is up and east is left in all images.}}
\label{fig:1705_1808}
\end{figure}

\addtocounter{figure}{-1}
\addtocounter{subfigure}{1}
\begin{figure}[htbp]
\centering
\includegraphics[width=1.0\textwidth]{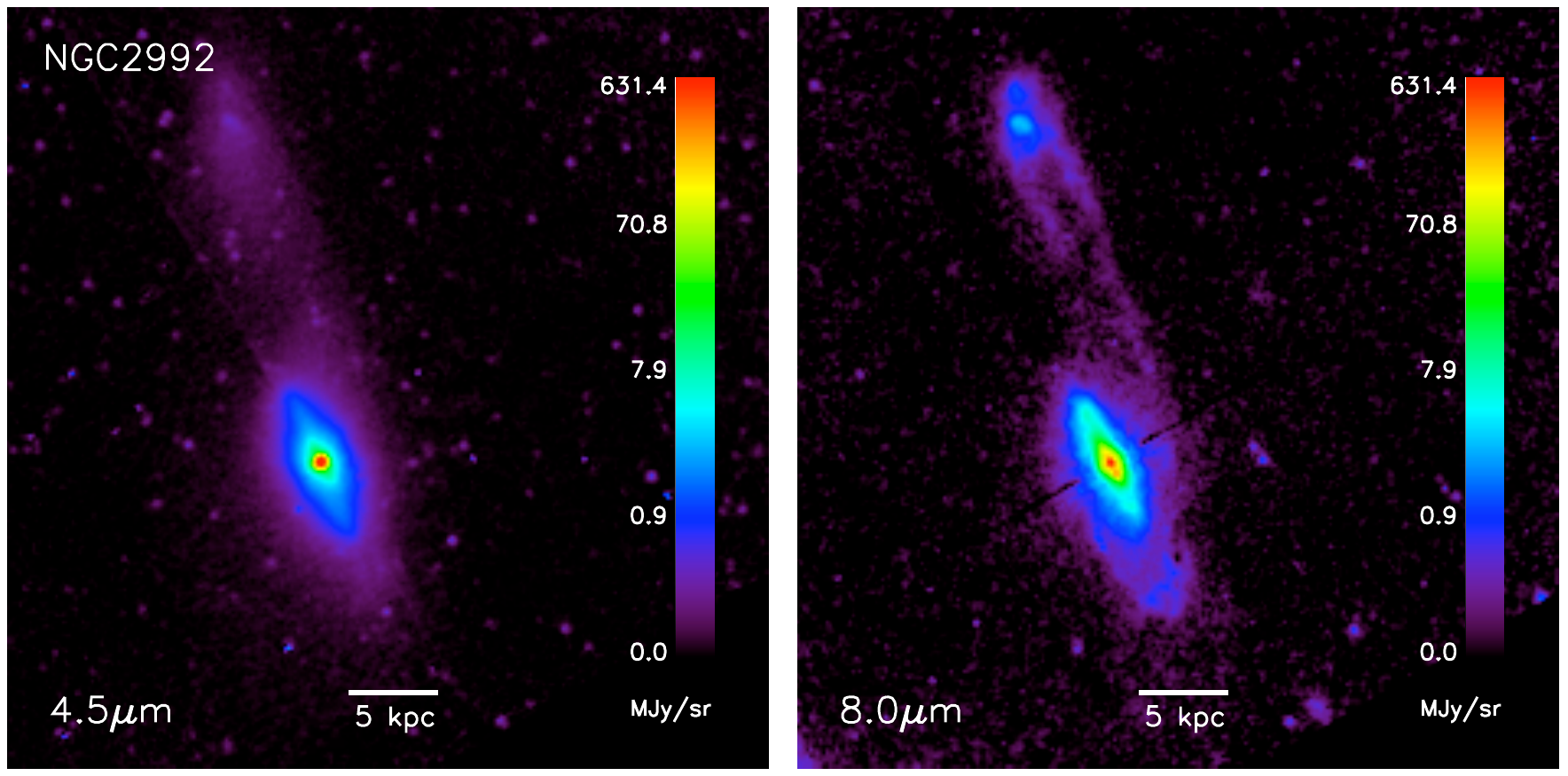}
\includegraphics[width=1.0\textwidth]{spacer}
\includegraphics[width=1.0\textwidth]{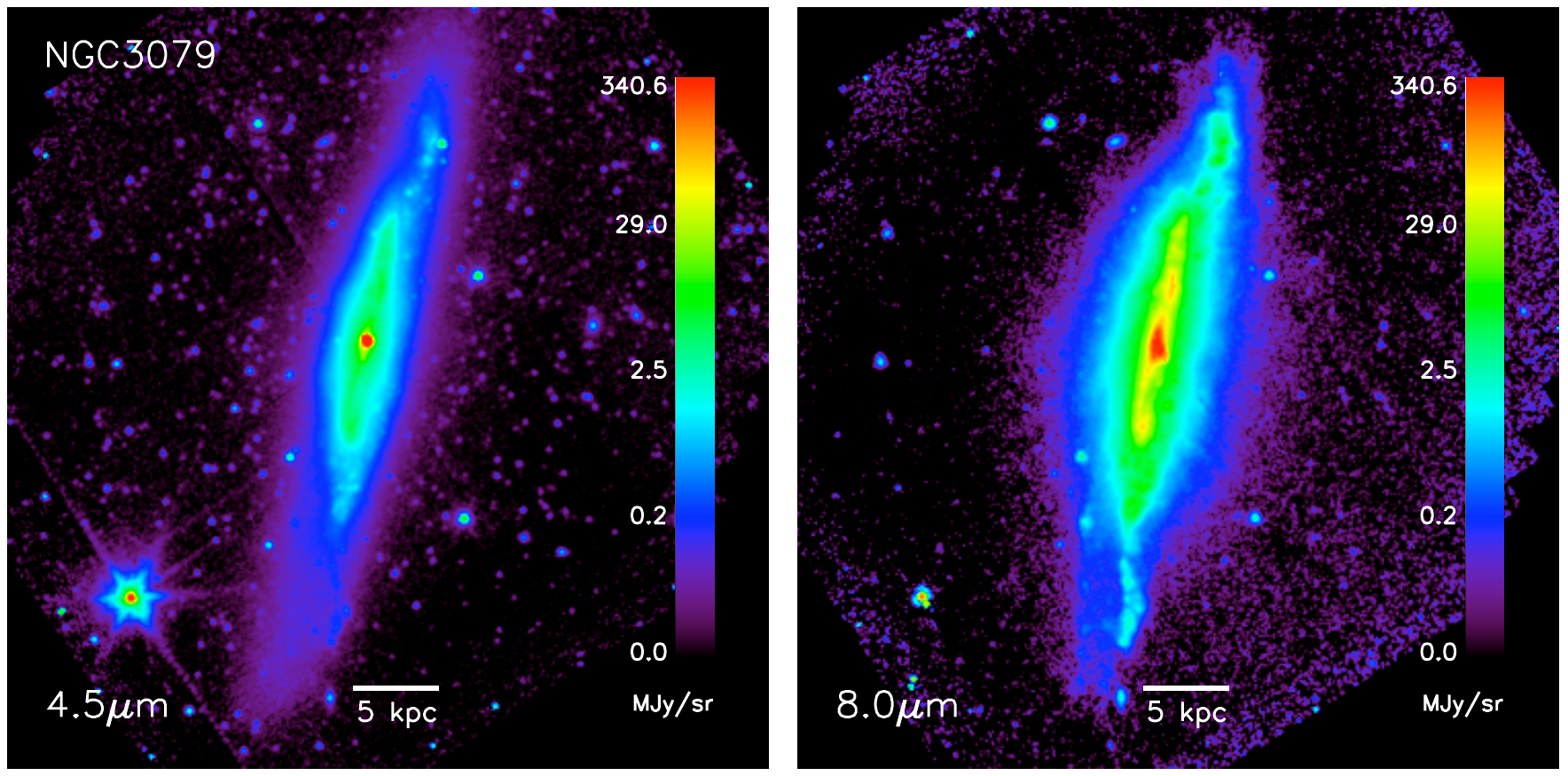}
\caption{\footnotesize{Comparing IRAC 4.5 and 8.0 $\mu$m maps of NGC 2992 and NGC 3079. The intensity scalings are "asinh" as described in Figure~\ref{fig:55}. North is up and east is left in all images.}}
\label{fig:2992_3079}
\end{figure}

\addtocounter{figure}{-1}
\addtocounter{subfigure}{1}
\begin{figure}[htbp]
\centering
\includegraphics[width=1.0\textwidth]{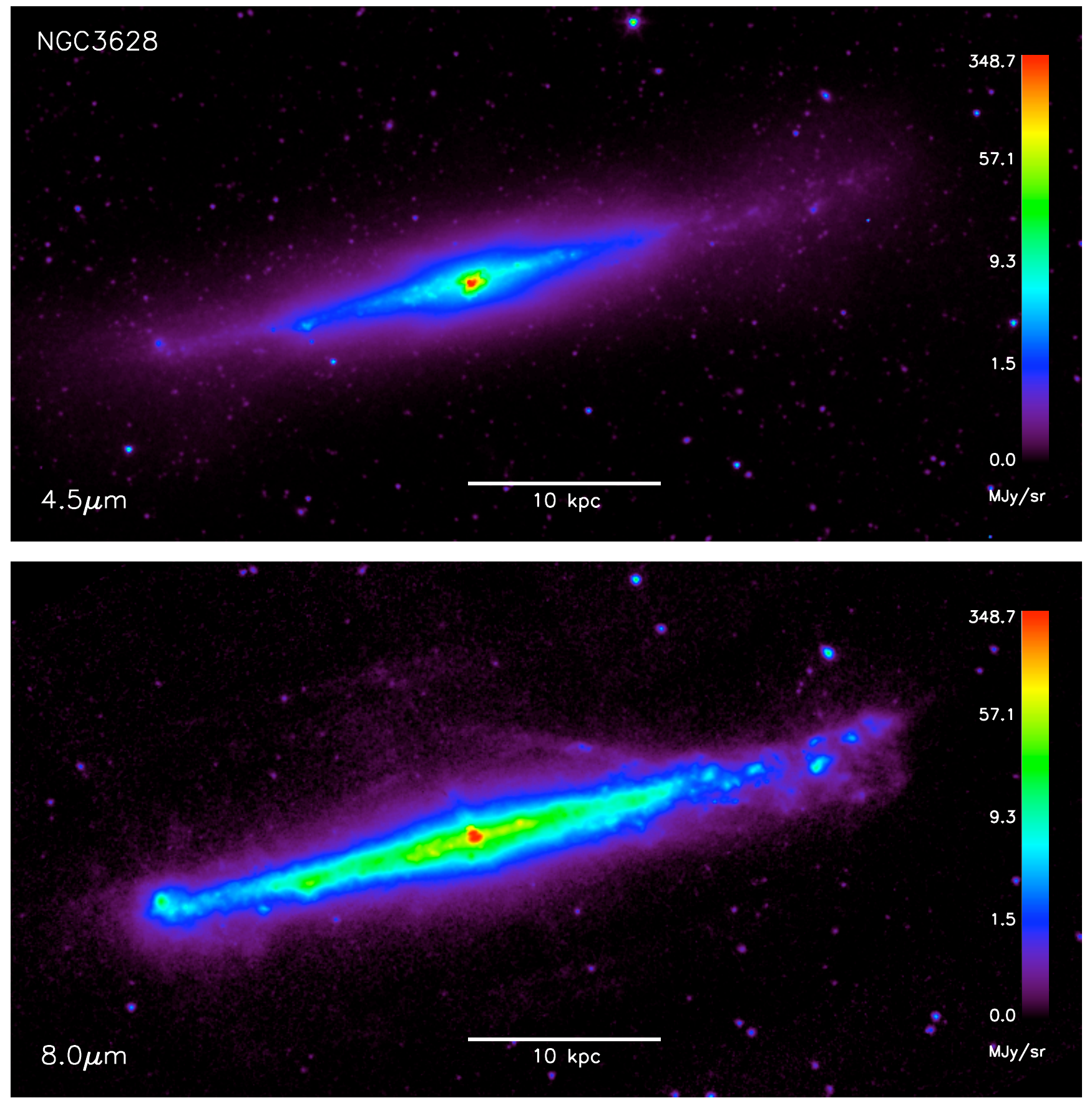}
\caption{\footnotesize{Comparing IRAC 4.5 and 8.0 $\mu$m maps of NGC 3628. The intensity scalings are "asinh" as described in Figure~\ref{fig:55}. North is up and east is left in both images.}}
\label{fig:3628}
\end{figure}

\addtocounter{figure}{-1}
\addtocounter{subfigure}{1}
\begin{figure}[htbp]
\centering
\includegraphics[width=1.0\textwidth]{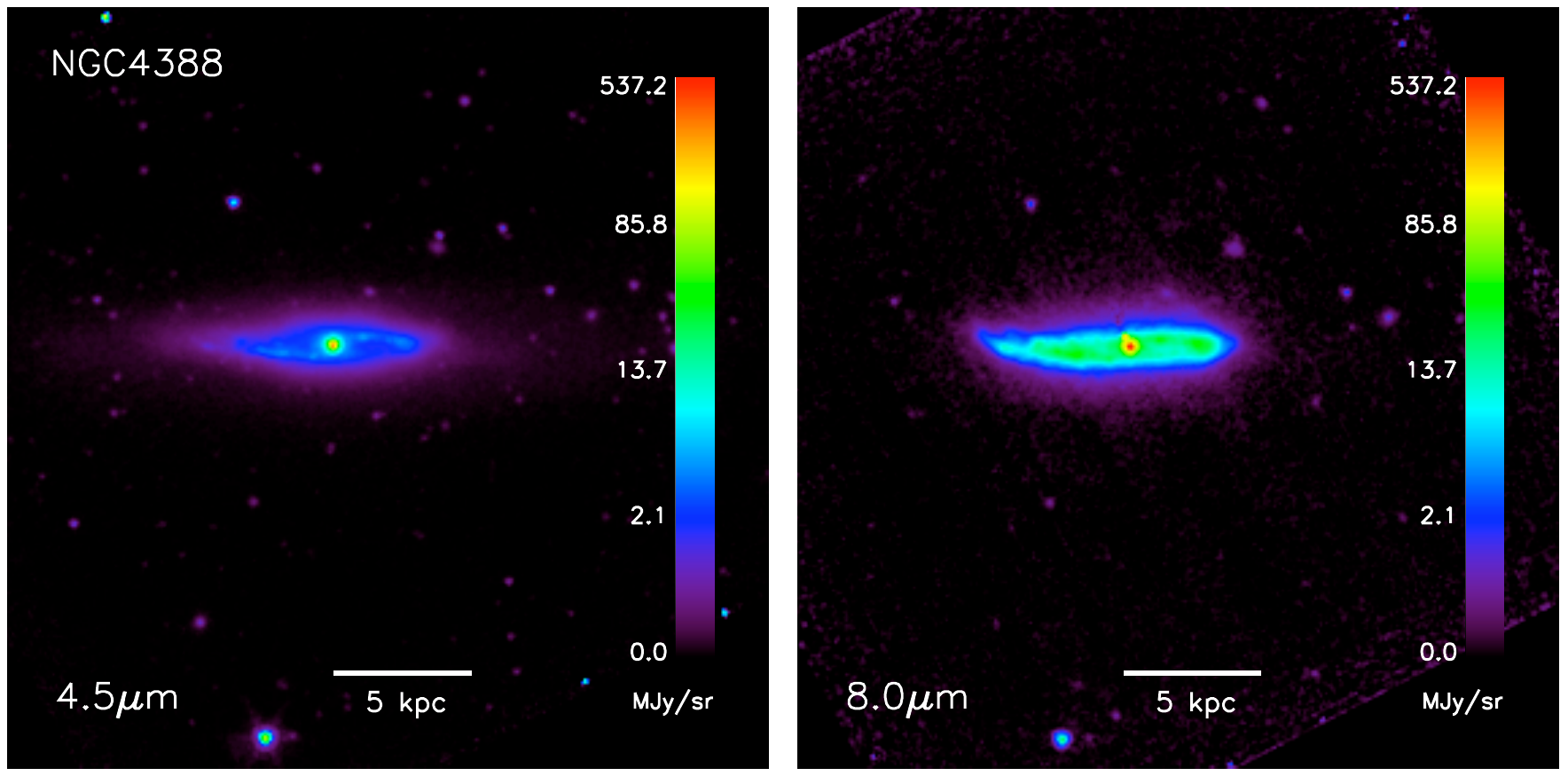}
\includegraphics[width=1.0\textwidth]{spacer}
\includegraphics[width=1.0\textwidth]{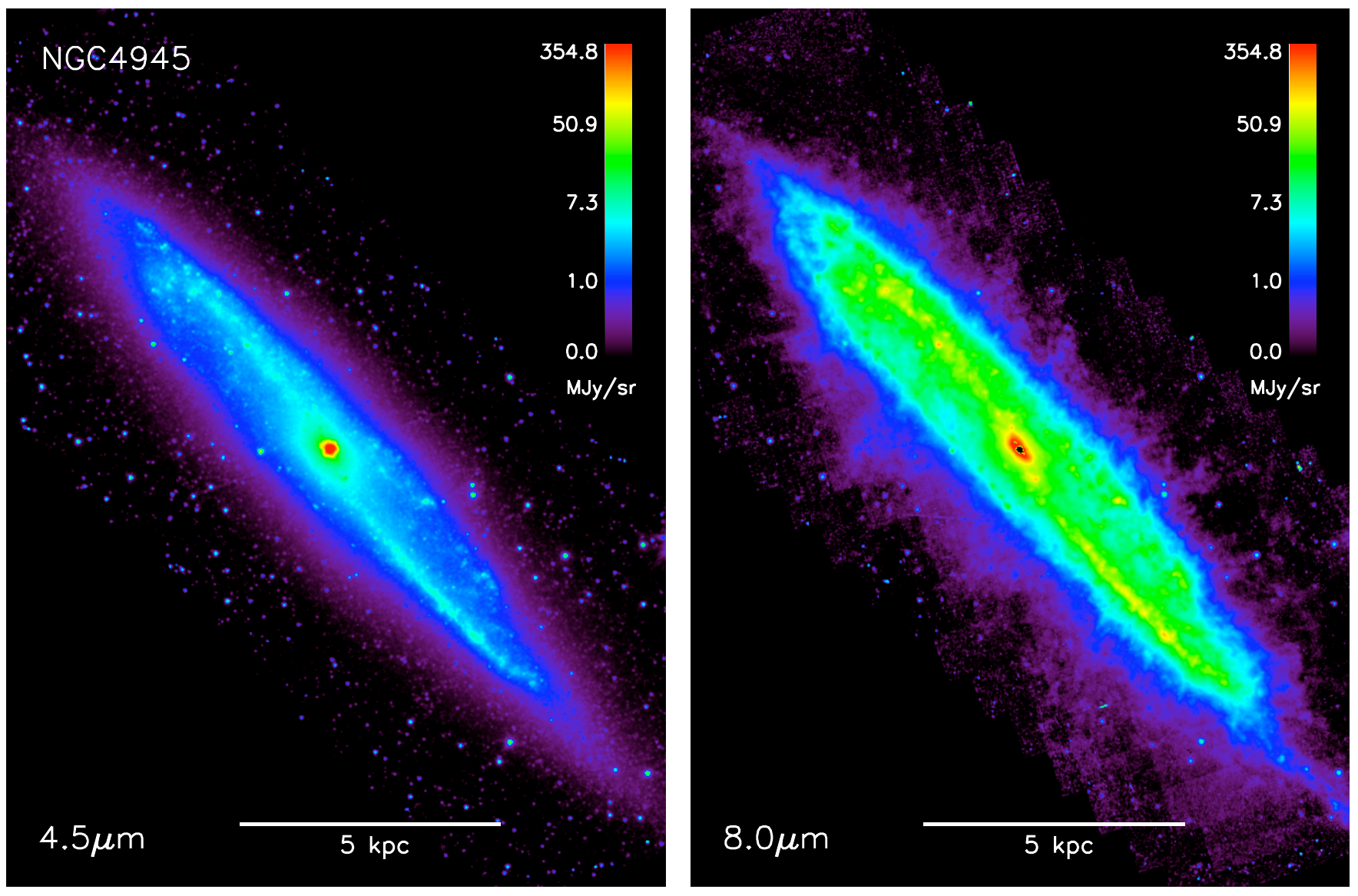}
\caption{\footnotesize{Comparing IRAC 4.5 and 8.0 $\mu$m maps of NGC 4388 and NGC 4945. The intensity scalings are "asinh" as described in Figure~\ref{fig:55}. North is up and east is left in all images.}}
\label{fig:4388_4945}
\end{figure}

\addtocounter{figure}{-1}
\addtocounter{subfigure}{1}
\begin{figure}[htbp]
\centering
\includegraphics[width=1.0\textwidth]{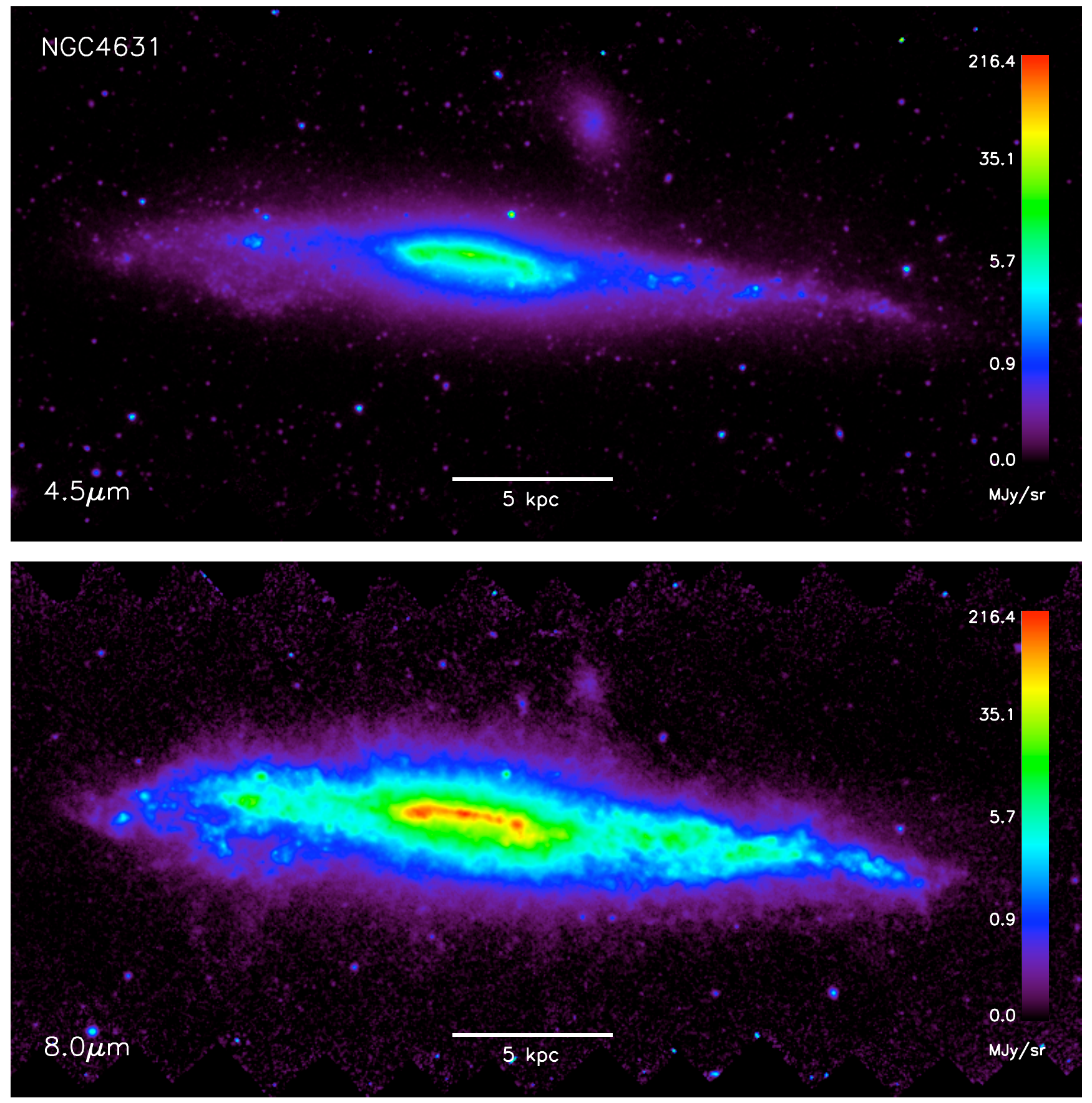}
\caption{\footnotesize{Comparing IRAC 4.5 and 8.0 $\mu$m maps of NGC 4631. The intensity scalings are "asinh" as described in Figure~\ref{fig:55}. North is up and east is left in both images.}}
\label{fig:4631}
\end{figure}

\addtocounter{figure}{-1}
\addtocounter{subfigure}{1}
\begin{figure}[htbp]
\centering
\includegraphics[width=1.0\textwidth]{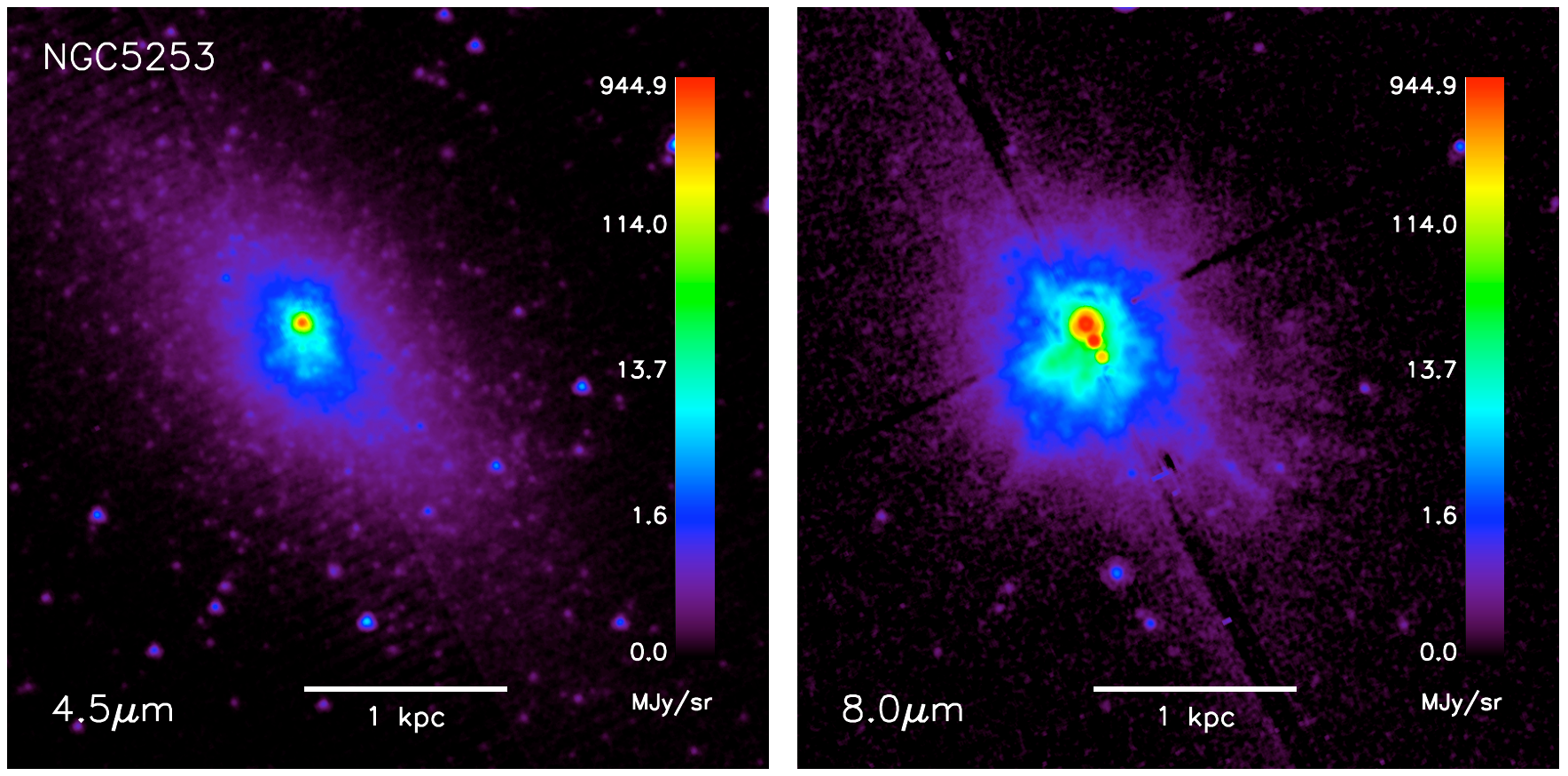}
\includegraphics[width=1.0\textwidth]{spacer}
\includegraphics[width=1.0\textwidth]{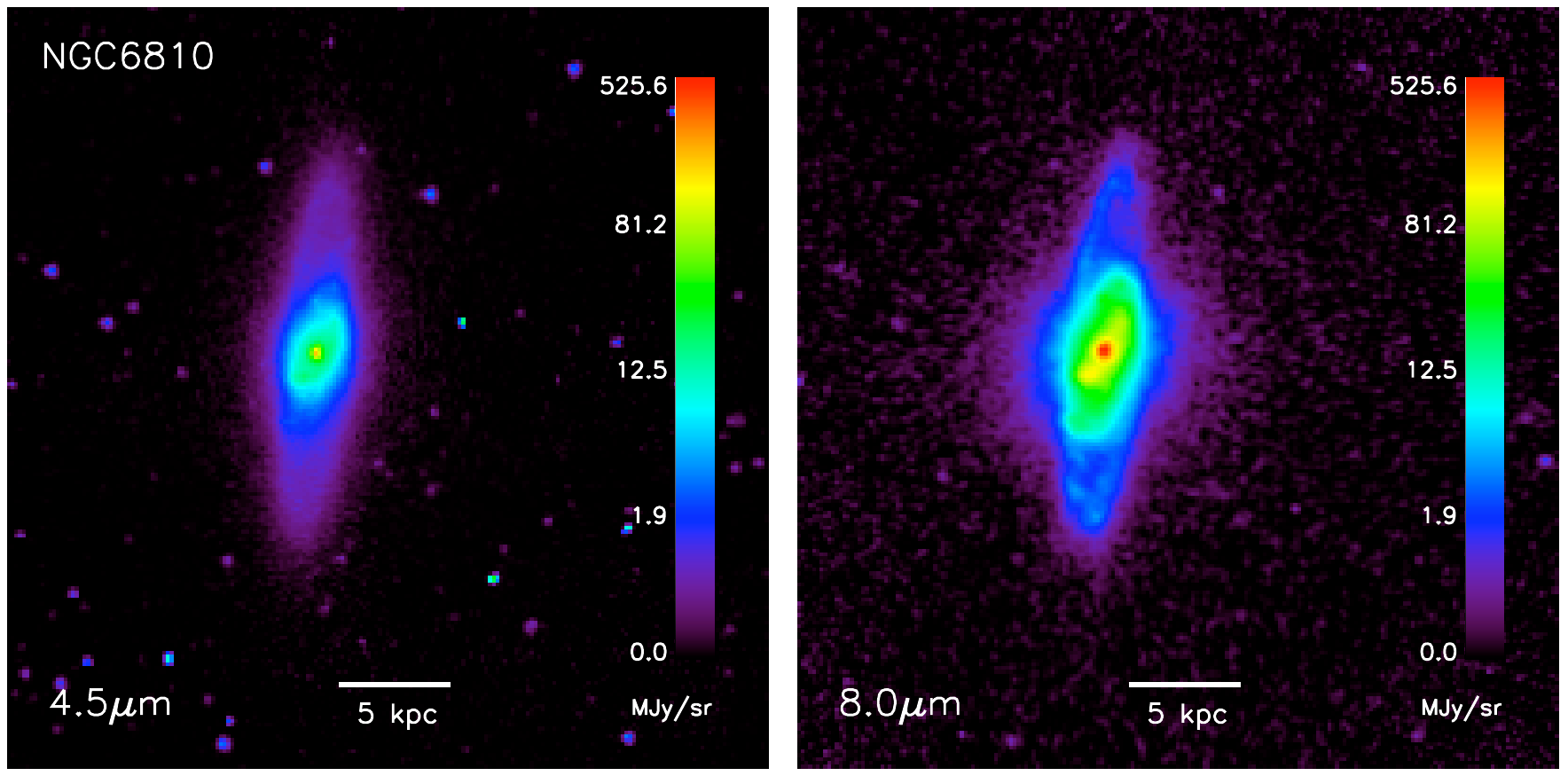}
\caption{\footnotesize{Comparing IRAC 4.5 and 8.0 $\mu$m maps of NGC 5253 and NGC 6810. The intensity scalings are "asinh" as described in Figure~\ref{fig:55}. North is up and east is left in all images.}}
\label{fig:5253_6810}
\end{figure}

\addtocounter{figure}{-1}
\addtocounter{subfigure}{1}
\begin{figure}[htbp]
\centering
\includegraphics[width=1.0\textwidth]{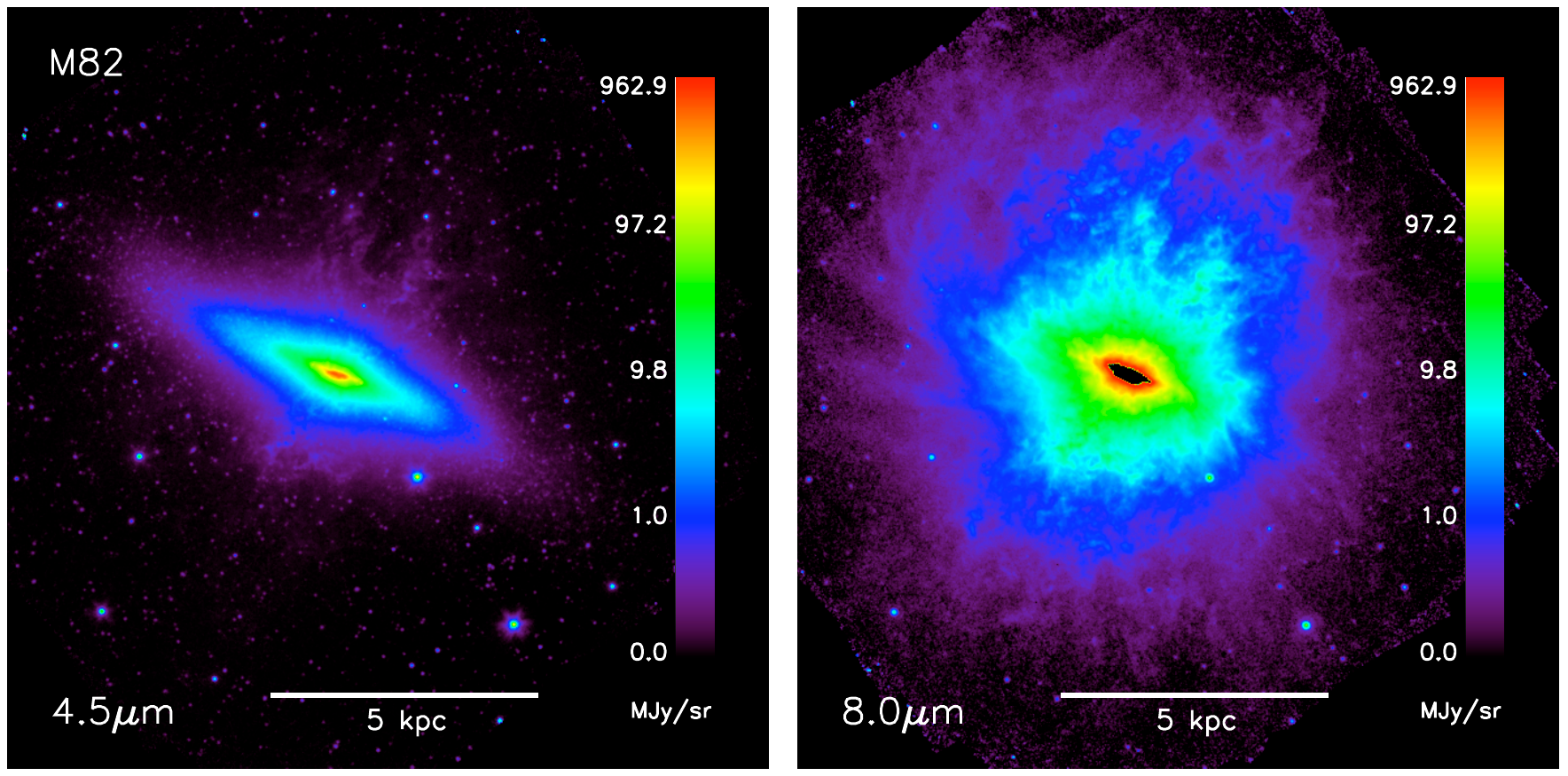}
\caption{\footnotesize{Comparing IRAC 4.5 and 8.0 $\mu$m maps of M82. The intensity scalings are "asinh" as described in Figure~\ref{fig:55}. North is up and east is left in both images.}}
\label{fig:82}
\end{figure}

\renewcommand{\thefigure}{\arabic{figure}}

\FloatBarrier

% PLOTS...

\begin{figure}[htbp]
\centering
\includegraphics[width=1.0\textwidth]{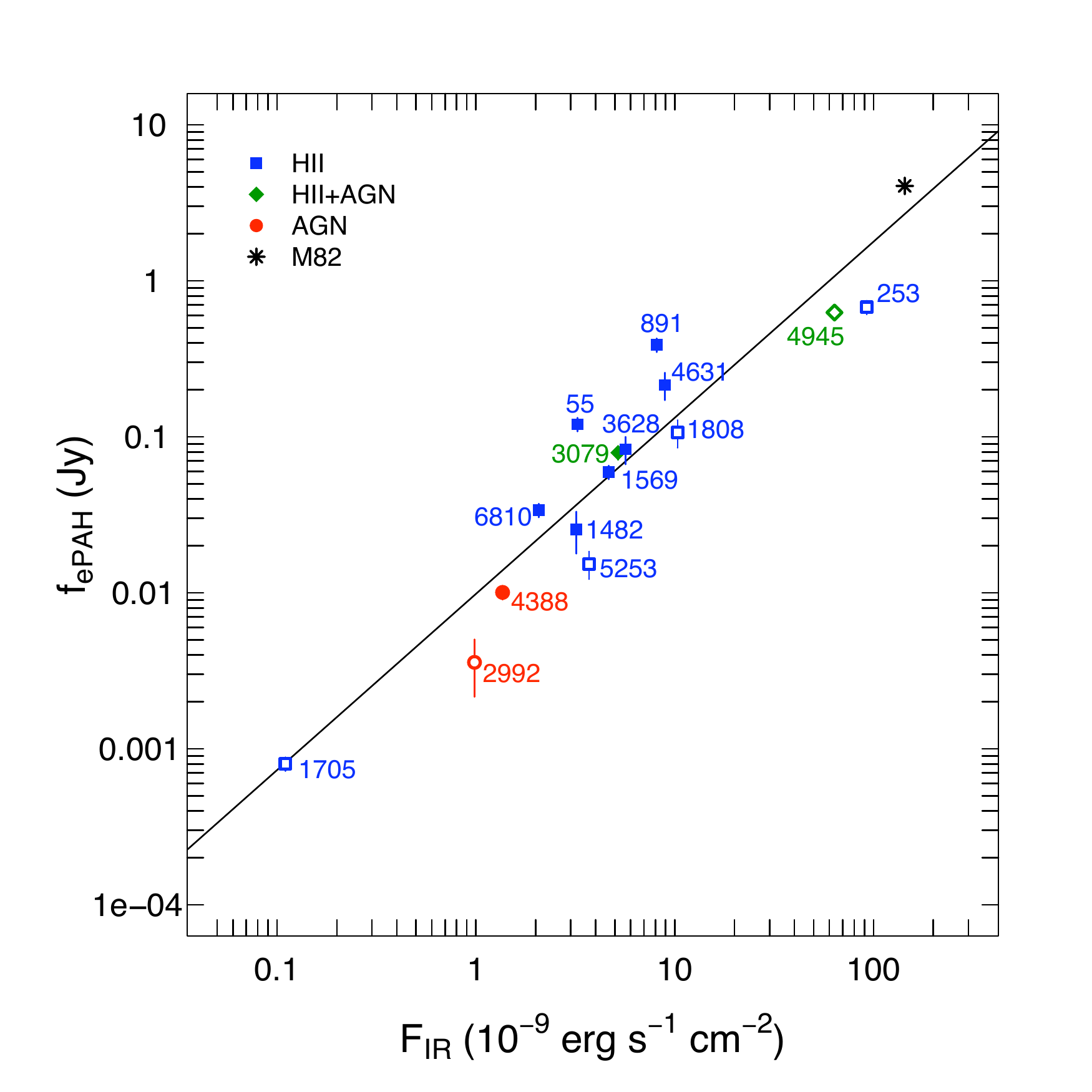}
\caption{\footnotesize{The extraplanar PAH flux, $f_{ePAH}$, plotted versus the calculated total IR flux, $F_{IR}$, on a log-log scale. Meaning of the symbols is the same as in Figure~\ref{fig:RossaDettmar_analog}. . The line is a power law fit with exponent 1.13 $\pm$ 0.11. (see \S~\ref{results_discuss}).}}
\label{fig:plot1}
\end{figure}

\begin{figure}[htbp]
\centering
\includegraphics[height=1.1\textwidth]{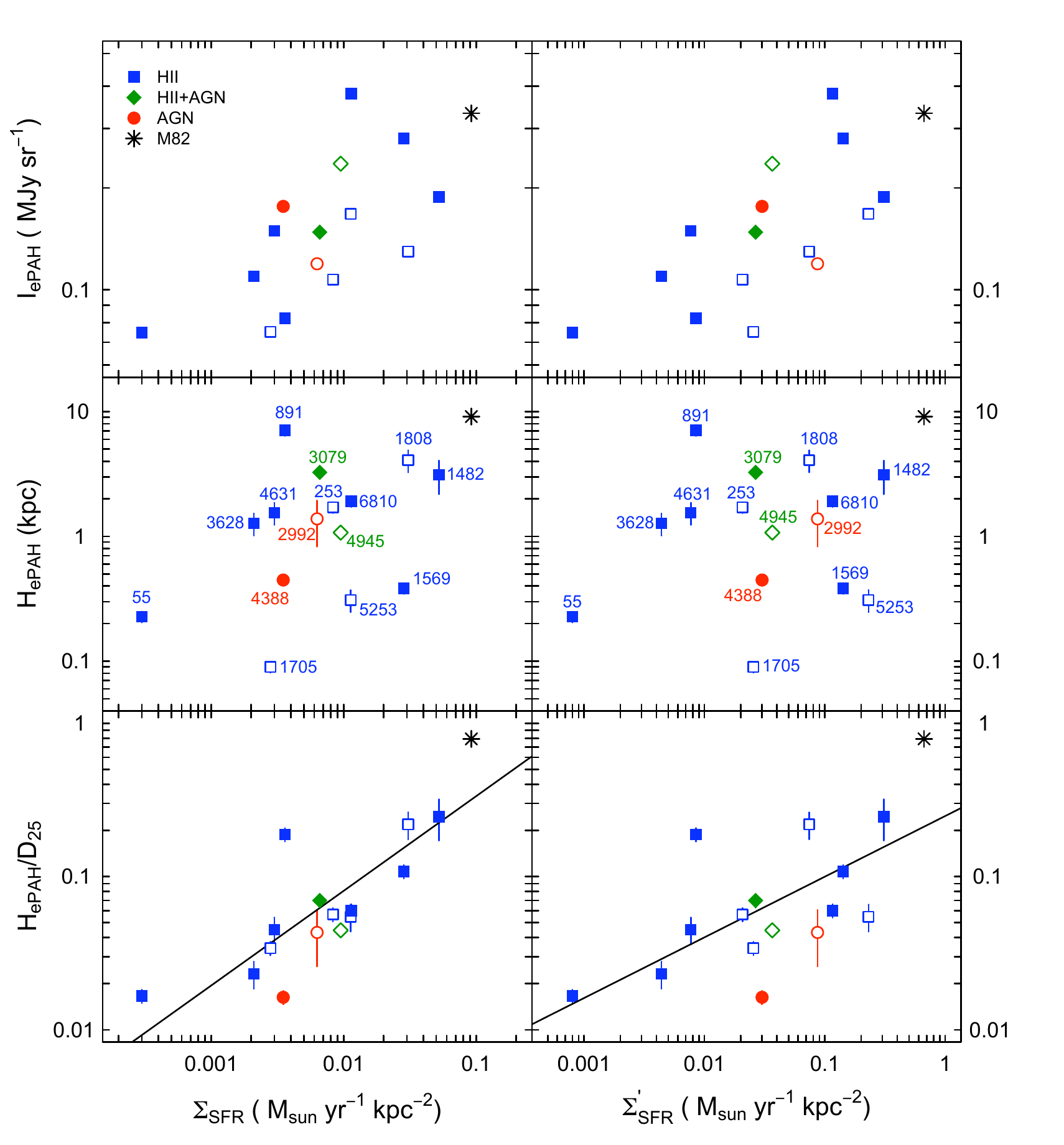}
\caption{\footnotesize{Characteristic extraplanar PAH surface brightness ($I_{ePAH}$), characteristic extraplanar emission height ($H_{ePAH}$), and $H_{ePAH}$ scaled by the stellar disk diameter ($H_{ePAH} / D_{25}$) plotted versus the galaxy's star formation rate (SFR) surface density ($\Sigma_{SFR}$) and characteristic SFR surface density ($\Sigma^{'}_{SFR}$, see \S~\ref{results_discuss}) on log-log scales. The best fit power laws shown in the bottom-left and bottom-right panels as solid lines have exponents of 0.62 $\pm$ 0.12 and 0.40 $\pm$ 0.12, respectively. Meaning of the symbols is the same as in Figure~\ref{fig:RossaDettmar_analog} (see \S~\ref{flux}).}}
\label{fig:plot2}
\end{figure}

% ----- TABLES -----

% SAMPLE
\begin{deluxetable}{l l l r c c c l}
\tabletypesize{\scriptsize}
\tablecolumns{8}
\tablewidth{0pt}
\tablecaption{Nearby Wind Galaxies from the {\em Spitzer} Archive and Their Properties \label{tbl:sample}}
\tablehead{
	\colhead{Galaxy} & 
	\colhead{Type} & 
	\colhead{Morph.} &
	\colhead{$D_{25}$ (')} & 
	\colhead{$i$ ($^\circ$)} & 
	\colhead{$d$ (Mpc)} & 
	\colhead{$L_{IR}$ (10$^{10}$$L_{\odot}$)} &
	\colhead{Wind/eDIG Ref(s).} \\
	\colhead{} &
	\colhead{[1]} &
	\colhead{[2]} &
	\colhead{[3]} &
	\colhead{[4]} &
	\colhead{[5]} &
	\colhead{[6]} &
	\colhead{[7]}	
}
\startdata
NGC 55 		& H {\sc ii} 		& SB(s)m 			& 29.3		& 80			& 1.6 		& 0.026 	& 1,2,3,4,5,6,7\\
NGC 253 		& H {\sc ii}			& SAB(s)c 		& 30.0		& 76			& 3.46 		& 3.44 	& 1,4,7,8,9,10,11,12,13,14,15\\
NGC 891 		& H {\sc ii}			& SA(s)b? 		& 13.5 		& 79			& 9.6 		& 2.32 	& 5,7,14,15,16,17,18\\
NGC 1482 	& H {\sc ii}			& SA0$^{+}$ pec 	& 2.2 		& 56			& 19.6 		& 3.82 	& 7,12,14,15,19,20\\
NGC 1569 	& H {\sc ii}			& IBm 			& 3.63 		& 61			& 3.36 		& 0.16 	& 21,22,23,24\\ 
NGC 1705 	& H {\sc ii}			& SA0$^{-}$ pec 	& 1.8 		& 42			& 5.1 		& 0.009 	& 25,26,27,28\\
NGC 1808 	& H {\sc ii}			& (R)SAB(s)a 		& 5.2 		& 53			& 12.3 		& 4.86 	& 12,29\\
NGC 2992 	& AGN 			& Sa pec 			& 3.5 		& 72			& 31 		& 2.94 	& 30,31,32,33\\
NGC 3079 	& H {\sc ii} + AGN	& SB(s)c 			& 7.9 		& 80			& 20.2 		& 6.58 	& 1,7,9,14,15,34,35,36\\
NGC 3628 	& H {\sc ii}			& Sb pec 			& 14.8 		& 78			& 12.7 		& 2.84 	& 1,7,12,14,15,37,38\\
NGC 4388 	& AGN 			& SA(s)b 			& 5.6 		& 77			& 16.7 		& 1.18 	& 30,31,32,39\\
NGC 4631 	& H {\sc ii}			& SB(s)d 			& 15.5 		& 80			& 7.62 		& 1.61 	& 1,7,8,14,15,40\\
NGC 4945 	& H {\sc ii} + AGN	& SB(s)cd 		& 23.3 		& 79			& 3.55 		& 2.49 	& 7,9,14,15,41,42\\
NGC 5253 	& H {\sc ii}			& pec 			& 5.2 		& 67			& 3.77 		& 1.64 	& 43\\
NGC 6810 	& H {\sc ii}			& SA(s)ab 		& 3.8 		& 74			& 28.7 		& 5.30 	& 12,18,44\\
M82			& H {\sc ii}			& I0				& 11.2		& 67			& 3.53		& 5.56	& 1,7,8,9,45,46,47\\
\enddata
\tablenotetext{1}{Optical/UV types are galaxies identified by star-forming H {\sc ii} regions (H {\sc ii}), active galactic nuclei (AGN), and galaxies containing both an AGN and star-forming H {\sc ii} regions (H {\sc ii} + AGN).}
\tablenotetext{2}{de Vaucouleurs morphological type from \cite{dev91} (hereafter RC3).}
\tablenotetext{3}{Diameter (major axis) in arcminutes based on 25th magnitude B-band observations (RC3 or \cite{lau89}).}
\tablenotetext{4}{Inclination in degrees with respect to the line of sight from the calculation of $\cos^{-1}(b/a)$ (RC3)}
\tablenotetext{5}{z-independent distance (except NGC 2992). References: NGC 55, NGC 891, and NGC 1482: \citealt{tul88}; NGC 253: \citealt{dal09}; NGC 1569: \citealt{gro08}; NGC 1705: \citealt{tos01}; NGC 1808 and NGC 4945: \citealt{tul09}; NGC 2992: \citealt{vei01}; NGC 3079 and NGC 6810: \citealt{spr09}; NGC 3628: \citealt{wil97}; NGC 4388: \citealt{yas97}; NGC 4631: \citealt{set05}; NGC 5253: \citealt{sak04}; M82: \citealt{kar02}}
\tablenotetext{6}{IR luminosity (8 - 1000 $\mu$m) expressed in units of 10$^{10}$$L_{\odot}$, calculated using equations in Table 1 of \cite{san96}, IRAS flux densities listed in the NASA/IPAC Extragalactic Database \citep{mos90,san03}, and the distances listed in this table.}
\tablenotetext{7}{Selected references to a galactic wind or extraplanar diffuse ionized gas (eDIG): (1) \citealt{dah98}; (2) \citealt{fer96}; (3) \citealt{gra82}; (4) \citealt{hoo96}; (5) \citealt{mil03}; (6) \citealt{ott99}; (7) \citealt{tul06}; (8) \citealt{alt99}; (9) \citealt{hec90}; (10) \citealt{hoo05}; (11) \citealt{pie00}; (12) \citealt{sha10}; (13) \citealt{str02}; (14) \citealt{str04a}; (15) \citealt{str04b}; (16) \citealt{alt98}; (17) \citealt{how97}; (18) \citealt{bre13}; (19) \citealt{ham99}; (20) \citealt{vei02}; (21) \citealt{del96}; (22) \citealt{hec95}; (23) \citealt{hun93}; (24) \citealt{wal91}; (25) \citealt{hec01}; (26) \citealt{meu89}; (27) \citealt{meu92}; (28) \citealt{meu98}; (29) \citealt{phi93}; (30) \citealt{col96}; (31) \citealt{col98}; (32) \citealt{gal06}; (33) \citealt{vei01}; (34) \citealt{cec01}; (35) \citealt{pie98}; (36) \citealt{vei94}; (37) \citealt{dah96}; (38) \citealt{fab90}; (39) \citealt{vei99}; (40) \citealt{wan01}; (41) \citealt{moo96}; (42) \citealt{nak89}; (43) \citealt{mar95}; (44) \citealt{str07}; (45) \citealt{eng06}; (46) \citealt{str97}; (47) \citealt{vei09}.}
\end{deluxetable}

% DATA SUMMARY
\begin{deluxetable}{l c c c l c c}
\tablecaption{{\em Spitzer} Archival Data \label{tbl:data}}
\tabletypesize{\scriptsize}
\tablecolumns{7}
\tablewidth{0pt}
\tablehead{
	\colhead{Galaxy} & 
	\colhead{Exp.} 			& 
	\colhead{$t_{exp}$ (s)} 	&
	\colhead{{\em Spitzer} AOR} 	&
	\colhead{PI (survey)}	& 
	\colhead{$\sigma_{4.5}$ (MJy sr$^{-1}$)} &
	\colhead{$\sigma_{8.0}$ (MJy sr$^{-1}$)} \\
	\colhead{}				&
	\colhead{[1]} 	&
	\colhead{[2]} 	&
	\colhead{[3]} 	&
	\colhead{[4]}	&
	\colhead{[5]} 	&
	\colhead{[6]}
}
\startdata
NGC 55 		& 180 	& 30 (26.8) 		& 14478592 	& de Jong, R. S. 	& 0.0062 & 0.0276 \\
\hline
NGC 253		& 135 	& 30 (26.8) 		& 18377728 	& Armus, L. 		& 0.0076 & 0.0252 \\
{}			& 72 	& 30 (26.8) 		& 18377984 	& {} & {} & {} \\
{}			& 168 	& 30 (26.8) 		& 18378240 	& {} & {} & {} \\
{}			& 168 	& 30 (26.8) 		& 18378496 	& {} & {} & {} \\
{}			& 48 	& 30 (26.8) 		& 18378752 	& {} & {} & {} \\
\hline
NGC 891		& 96 	& 100 (96.8) 		& 3631872 	& Fazio, G. 		& 0.0171 & 0.0130 \\
{}			& 128 	& 100 (96.8) 		& 3632128 	& {} & {} & {} \\
{}			& 128 	& 100 (96.8) 		& 3632384 	& {} & {} & {} \\
\hline
NGC 1482 	& 4 		& 30 (26.8) 		& 5533696 	& Kennicutt, R. (SINGS) & 0.0104 & 0.0450 \\
\hline
NGC 1569 	& 5 		& 12 (10.4) 		& 4434944 	& Fazio, G.		& 0.0220 & 0.0850 \\ 
\hline
NGC 1705	& 4 		& 30 (26.8) 		& 5535744 	& Kennicutt, R. (SINGS) & 0.0068 & 0.0303 \\
{}			& 4 		& 30 (26.8) 		& 5536000 	& {} & {} & {} \\
\hline
NGC 1808 	& 16 	& 30 (26.8) 		& 18284800 	& Fisher, D. B. 		& 0.0061 & 0.0264 \\
\hline
NGC 2992 	& 15 	& 12 (10.4) 		& 4933376 	& Houck, J. R. 		& 0.0139 & 0.0425 \\
\hline
NGC 3079 	& 16 	& 30 (26.8) 		& 22000896 	& Rieke, G. 		& 0.0061 & 0.0292 \\
\hline
NGC 3628 	& 36 	& 30 (26.8) 		& 22547200 	& Kennicutt, R. (LVL) & 0.0082 & 0.0378 \\
{}			& 36 	& 30 (26.8) 		& 22547456 	& {} & {} & {} \\
\hline
NGC 4388 	& 9 		& 12 (10.4) 		& 27090688 	& Forman, W. R. 	& 0.0134 & 0.0598 \\
\hline
NGC 4631 	& 15 	& 30 (26.8) 		& 5540864 	& Kennicutt, R. (SINGS) & 0.0081 & 0.0359 \\
{}			& 15 	& 30 (26.8) 		& 5541120 	& {} & {} & {} \\
\hline
NGC 4945 	& 30 	& 30 (26.8) 		& 21999616 	& Rieke, G. 		& 0.0142 & 0.0636 \\
\hline
NGC 5253 	& 24 	& 12 (10.4) 		& 4386048 	& Houck, J. R. 		& 0.0131 & 0.0466 \\
\hline
NGC 6810 	& 1 		& 12 (10.4) 		& 12471808 	& Gallimore, J. F. 	& 0.0583 & 0.1331 \\
\hline
M82			& 10		& 12 (10.4)		& 13457920	& Kennicutt, R. (SINGS) & 0.0081 & 0.0443 \\
{}			& 20		& 30 (26.8)		& 13458176	& {} & {} & {} \\
{}			& 20		& 30 (26.8)		& 13458432	& {} & {} & {} \\
{}			& 10		& 12 (10.4)		& 13459200	& {} & {} & {} \\
\enddata
\tablenotetext{1}{\scriptsize Exp. = number of exposures used for our reduction - many extraneous exposures were excluded. The same number of exposures were used from each IRAC channel.}
\tablenotetext{2}{\scriptsize Integration time for each frame (effective integration time per pixel) in seconds.}
\tablenotetext{3}{\scriptsize {\em Spitzer} Astronomical Observation Request identifing the data set we used for each galaxy in the {\em Spitzer} archive.}
\tablenotetext{4}{\scriptsize Principal Investigator and survey (e.g. SINGS), where applicable.}
\tablenotetext{5,6}{\scriptsize $\sigma$ of the background noise for the combined/mosaicked IRAC 4.5 and 8.0 $\mu$m images.}
\end{deluxetable}

% RESULTS
\begin{deluxetable}{l c c c c c c c c c c c c c c}
\tabletypesize{\scriptsize}
\tablewidth{0pt}
\tablecolumns{12}
\tablecaption{IR Luminosity and Extraplanar PAH Emission \label{tbl:results}}
\tablehead{
	\colhead{Galaxy} &
	\colhead{Type} &
	\colhead{$F_{IR}$} &
	\colhead{$L_{IR}$} &
	\colhead{$f_{ePAH}$} &
	\colhead{$D_{4.5 \mu m}$} &
	\colhead{$I_{ePAH}$} &
	\colhead{$H_{ePAH}$} &
	\colhead{$\Sigma_{SFR}$} &
	\colhead{$D_{PAH}$} &
	\colhead{$\Sigma^{'}_{SFR}$} &
	\colhead{$z_{ext}$} \\
	\colhead{} &
	\colhead{} &
	\colhead{[1]} 	&
	\colhead{[2]} 	&
	\colhead{[3]} 	&
	\colhead{[4]} 	&
	\colhead{[5]} 	&
	\colhead{[6]} 	&
	\colhead{[7]} 	&
	\colhead{[8]} 	&
	\colhead{[9]} 	&
	\colhead{[10]} 
}
\startdata
NGC 55 		& H {\sc ii} 		& 3.24	& 0.026 	& 0.120  	& 9.3 	& 0.075	& 0.227	& 0.0003	& 8.3	& 0.0008	& 1.0 \\
NGC 253 		& H {\sc ii}			& 92.5 	& 3.440 	& 0.677   	& 22.2 	& 0.107	& 1.709	& 0.0083	& 19.1	& 0.0207	& 5.2 \\
NGC 891 		& H {\sc ii}			& 8.12	& 2.324 	& 0.388   	& 29.2 	& 0.082	& 7.095	& 0.0036	& 24.5	& 0.0085	& 6.0 \\
NGC 1482 	& H {\sc ii}			& 3.20 	& 3.822 	& 0.025 :  	& 6.7 	& 0.188	& 3.113 :	& 0.0523	& 5.2	& 0.3083	& 4.1 \\
NGC 1569 	& H {\sc ii}			& 4.66 	& 0.163 	& 0.059   	& 2.5 	& 0.280	& 0.384	& 0.0285	& 1.6	& 0.1420	& 0.8 \\
NGC 1705 	& H {\sc ii}			& 0.110 	& 0.009 	& 0.001   	& 1.0 	& 0.075	& 0.090	& 0.0028	& 0.9	& 0.0255	& \nodata \\
NGC 1808 	& H {\sc ii}			& 10.3 	& 4.863 	& 0.106 :  	& 14.8 	& 0.130	& 4.079 :	& 0.0307	& 12.0	& 0.0741	& 6.0 \\
NGC 2992 	& AGN	 		& 0.984 	& 2.939 	& 0.004 :  	& 12.3 	& 0.119	& 1.382 :	& 0.0063	& 8.6	& 0.0869	&  3.9 \\
NGC 3079 	& H {\sc ii} + AGN	& 5.19 	& 6.580 	& 0.079   	& 28.0 	& 0.148	& 3.260	& 0.0066	& 23.3	& 0.0266	&  5.5 \\
NGC 3628 	& H {\sc ii}			& 5.66 	& 2.836 	& 0.083 : 	& 38.7 	& 0.110	& 1.270 :	& 0.0021	& 37.8	& 0.0044	&  5.0 \\
NGC 4388 	& AGN			& 1.36 	& 1.181 	& 0.010   	& 13.9 	& 0.176	& 0.447	& 0.0035	& 9.3	& 0.0301	&  2.1 \\
NGC 4631 	& H {\sc ii}			& 8.93 	& 1.612 	& 0.215 :  	& 24.1 	& 0.149	& 1.547 :	& 0.0030	& 21.5	& 0.0077	&  3.8 \\
NGC 4945 	& H {\sc ii} + AGN	& 63.6 	& 2.493 	& 0.626  	& 16.7 	& 0.236	& 1.072	& 0.0095	& 12.2	& 0.0366	& 2.9 \\
NGC 5253 	& H {\sc ii}			& 3.71 	& 0.164 	& 0.015 :  	& 2.6 	& 0.167	& 0.309 :	& 0.0113	& 1.2	& 0.2306	& 0.9 \\
NGC 6810 	& H {\sc ii}			& 2.07 	& 5.299 	& 0.034  	& 19.0 	& 0.380	& 1.912	& 0.0114	& 10.0	& 0.1160	&  5.3 \\
M82			& H {\sc ii}			& 144	& 5.563 	& 4.055	& 8.0 	& 0.333	& 9.120	& 0.0920	& 4.3	& 0.6666	&  6.0 \\
\enddata
\tablenotetext{1,2}{The IR flux [10$^{-9}$ erg s$^{-1}$ cm$^{-2}$] and luminosity [10$^{10}$ $L_{\odot}$] calculated using equations in Table 1 of \cite{san96}, IRAS flux densities listed in the NASA/IPAC Extragalactic Database \citep{mos90,san03}, and the distances listed in Table~\ref{tbl:sample}.}
\tablenotetext{3}{Extraplanar 8.0 $\mu$m PAH flux [Jy] measured as described in \S~\ref{flux}. The uncertainties for the flux values are within $\sim$10-20\% (systematic photometric uncertainty) and values followed by a : fall within $\sim$20-40\% (systematic photometric uncertainty plus persistent artifacts). Flux uncertainties also apply to the corresponding $H_{ePAH}$ values.}
\tablenotetext{4}{Major axis [kpc] in the IRAC 4.5 $\mu$m images at the 0.175 MJy sr$^{-1}$ contour.}
\tablenotetext{5}{Characteristic intensity of extraplanar PAH emission [MJy sr$^{-1}$], see \S~\ref{PAHproperties}.}
\tablenotetext{6}{Characteristic height of extraplanar PAH emission [kpc], see \S~\ref{PAHproperties}. Uncertainty percentages are the same as in 3.}
\tablenotetext{7}{SFR surface density [$M_{\odot}$ yr$^{-1}$ kpc$^{-2}$] calculated using $L_{IR}$ and $D_{25}$, see \S~\ref{PAHproperties}.}
\tablenotetext{8}{Characteristic diameter of PAH emission (8.0 $\mu$m) in the disk [kpc] used to parametrize the location and distribution of star formation.}
\tablenotetext{9}{Characteristic SFR surface density [$M_{\odot}$ yr$^{-1}$ kpc$^{-2}$] calculated using $L_{IR}$ and $D_{PAH}$, see \S~\ref{PAHproperties}.}
\tablenotetext{10}{The furthest extent [kpc] of any PAH features projected perpendicular to the galaxy's major axis, see \S~\ref{z_ext}.}
\end{deluxetable}

\end{document}